\documentclass[fleqn,usenatbib]{mnras}

\usepackage{newtxtext,newtxmath}

\usepackage[T1]{fontenc}

\DeclareRobustCommand{\VAN}[3]{#2}
\let\VANthebibliography\thebibliography
\def\thebibliography{\DeclareRobustCommand{\VAN}[3]{##3}\VANthebibliography}

\usepackage{graphicx}	
\usepackage{amsmath}	

\title[Atmospheric Transmission Spectrum of WASP-17b]{A Comprehensive Analysis of WASP-17b's Transmission Spectrum from Space-Based Observations}

\author[L. Alderson et al.]{L. Alderson$^{1}$\thanks{E-mail: lili.alderson@bristol.ac.uk (LA)},
H. R. Wakeford$^{1}$, 
R. J. MacDonald$^{2}$, 
N. K. Lewis$^{2}$, 
E. M. May$^{3}$,
\newauthor D. Grant$^{1}$,
D. K. Sing$^{4,5}$,
K. B. Stevenson$^{3}$, 
J. Fowler$^{6}$, 
J. Goyal$^{7}$, 
N. E. Batalha$^{8}$, 
\newauthor  T. Kataria$^{9}$
\\
$^{1}$School of Physics, University of Bristol, HH Wills Physics Laboratory, Tyndall Avenue, Bristol BS8 1TL, UK\\
$^{2}$ Department of Astronomy and Carl Sagan Institute, Cornell University 122 Sciences Drive, Ithaca, NY, 14853, USA \\ 
$^{3}$ Johns Hopkins APL, 11100 Johns Hopkins Rd, Laurel, MD 20723, USA \\ 
$^{4}$ Department of Earth {\rm \&} Planetary Sciences, Johns Hopkins University, Baltimore, MD, USA\\ 
$^{5}$ Department of Physics {\rm \&} Astronomy, Johns Hopkins University, Baltimore, MD, USA\\
$^{6}$ Department of Astronomy and Astrophysics, University of California, Santa Cruz, CA 95064, USA \\ 
$^{7}$ School of Earth {\rm \&} Planetary Sciences (SEPS), National Institute of Science Education {\rm \&} Research (NISER), Bhubaneshwar - Odisha 752050, India \\
$^{8}$ NASA Ames Research Center, MS 245-3, Mountain View, CA 94035, USA \\
$^{9}$ NASA Jet Propulsion Laboratory, 4800 Oak Grove Dr, Pasadena, CA 91109, USA }

\date{Accepted 2022 March 3. Received 2022 March 3; in original form 2021 December 24}

\pubyear{2021}

\begin{document}
\label{firstpage}
\pagerange{\pageref{firstpage}--\pageref{lastpage}}
\maketitle

\begin{abstract}


Due to its 1770\,$K$ equilibrium temperature, WASP-17b, a 1.99\,$R_\mathrm{Jup}$, 0.486\,$M_\mathrm{Jup}$ exoplanet, sits at the critical juncture between hot and ultra-hot Jupiters. We present its 0.3--5\,$\micron$ transmission spectrum, with newly obtained with Hubble Space Telescope (HST) Wide Field Camera 3 (WFC3) measurements, and, taking advantage of improved analysis techniques, reanalysed HST Space Telescope Imaging Spectrograph (STIS) and Spitzer Space Telescope Infrared Array Camera (IRAC) observations. We achieve a median precision of 132\,ppm with a mean of 272\,ppm across the whole spectrum. We additionally make use of Transiting Exoplanet Survey Satellite (TESS) and ground-based transit observations to refine the orbital period of WASP-17b. To interpret the observed atmosphere, we make use of free and equilibrium chemistry retrievals using the \verb|POSEIDON| and \verb|ATMO| retrieval codes respectively. We detect absorption due to H$_2$O at $>7 \sigma$, and find evidence of absorption due to CO$_2$ at $>3 \sigma$. We see no evidence of previously detected \ion{Na}{i} and \ion{K}{i} absorption. Across an extensive suite of retrieval configurations, we find the data favours a bimodal solution with high or low metallicity modes, as a result of poor constraints in the optical and demonstrate the importance of using multiple statistics for model selection. Future James Webb Space Telescope (JWST) GTO observations, combined with the presented transmission spectrum, will enable precise constraints on WASP-17b’s atmosphere.

\end{abstract}




\section{Introduction} \label{section:info}

The atmospheres of giant exoplanets encode key information about the formation and evolution of these worlds beyond the solar system \citep[e.g.,][]{Oberg2011, Madhu2014, Mordasini2016, Espinoza2017, Eistrup2018}. In the past decade, the study of atmospheres of close-in gas giant exoplanets in transmission has been dominated by observations performed by the \textit{Hubble Space Telescope} \citep[HST, e.g.,][]{Charbonneau2002, Gibson2012, Huitson2012, Mandell2013, Nikolov2014, Kreidberg2015, Sing2015, Sing2016, Evans2016, Wakeford2017Sci, Spake2018, Wakeford2018, Alam2020, Lewis2020, Sheppard2021AJ}, characterising absorption by chemical species and scattering by aerosols. Many of these studies include observations with HST's Wide Field Camera 3 (WFC3), in particular making use of the instrument's near infra-red (NIR) capabilities via the G141 grism to measure prominent H$_2$O absorption features \citep[e.g.,][]{Deming2013, Fraine2014, Kreidberg2014_gj1214, Evans2016, Colon2020}. As a result, the 1.4\,$\micron$\, H$_2$O absorption feature has formed the backbone of many exoplanet comparative studies \citep[e.g.,][]{Iyer2016, Sing2016, Stevenson2016, Fu2017, Fisher2018, Tsiaras2018, Pinhas2019, Welbanks2019, Gao2020, Yu2021, Dymont2021}. However, NIR observations cannot alone precisely characterise the atmosphere of exoplanets, as the amplitudes and shapes of chemical absorption features are degenerate with the impacts of aerosol scattering and absorption towards the optical \citep[e.g.,][]{Fisher2018}. 

Accurately measuring the shape of the transmission spectrum into the optical towards ultra-violet (UV) wavelengths is therefore a vital part of understanding the nature of a planet's atmosphere \citep[e.g.,][]{Sing2011, McCullough2014, Pinhas2019, Bruno2020, Wakeford2020, Lewis2020}. Such observations, which capture information about absorption from dominant optical sources such as \ion{Na}{I}, \ion{K}{I}, TiO and VO, as well as wavelength dependent scattering towards the blue, can be performed by the Space Telescope Imaging Spectrograph (STIS) aboard HST \citep[e.g.,][]{Nikolov2015,Alam2021}, although they are also frequently made with ground-based telescopes, which provide the opportunity for continuous observations not afforded by HST \citep[e.g.,][]{Nikolov2016, Chen2018, Kirk2018, May2018, Alderson2020, Sotzen2020, Wilson2020, Weaver2021}. 

It is only with this combination of both optical and infra-red (IR) data that the complexities hidden within exoplanetary atmospheres can begin to be explored, as multiwavelength observations allow for multiple pressure depths and opacity sources to be probed. Such panchromatic studies, often complimented by photometry performed by the \textit{Spitzer Space Telescope} Infrared Array Camera (IRAC) at 3.6 and 4.5\,$\micron$ \citep[e.g.,][]{Sing2013, Wakeford2018, Alam2020, Carter2019, Spake2021}, have placed constraints on the abundances of key spectroscopically active species such as H$_2$O, Na and K \citep{Spake2021}, determined the role of uniform clouds and wavelength dependent scattering \citep{Sing2013, Alam2020}, and obtained the metallicities of atmospheres relative to solar values \citep{Wakeford2018, Carter2019}. The future of comprehensive space-based exoplanet atmosphere interpretation lies with the James Webb Space Telescope (JWST), GTO observations of which seek to obtain the 0.6--14$\micron$ spectrum of WASP-17b \citep{Anderson2010}. 

WASP-17b is a 1.991\,$R_\mathrm{Jup}$, 0.486\,$M_\mathrm{Jup}$, hot Jupiter, with a density of just 0.08\,$\rho_\mathrm{Jup}$ \citep{Anderson2011}, orbiting an 11.59 M$_V$, F6 type star in retrograde. With a period of 3.735\,d \citep{Triaud2010} and a 4.4-hour transit duration, WASP-17b is excellent target for transmission spectroscopy, aided by its large 1609\,km scale height (assuming a mean molecular weight for a H/He dominated atmosphere of 2.3). With an equilibrium temperature of 1770\,$K$ \citep{Anderson2011}, and a day- to night-side temperature range spanning 1000-2500\,$K$ \citep{Kataria2016}, WASP-17b is likely to play host to a broad range of spectroscopically active chemical species such as H$_2$O, CO$_2$ and CO from the optical to the near-IR \citep[e.g.,][]{Madhu2016}. At these temperatures, magnesium silicates are expected to be the dominant cloud species by both mass and opacity \citep[e.g.,][]{Visscher2010, Gao2020}. WASP-17b is an excellent probe of the parameter space between hot and ultra-hot Jupiters, such that the atmosphere of WASP-17b may be warm enough for TiO and VO to have a significant opacities \citep{Fortney2008}. 

As a result of the range of factors that make WASP-17b an excellent target, it was among the early vanguard of exoplanets to be probed via spectroscopic transmission observations. Ground-based high resolution atmospheric studies have found evidence of excess absorption caused by \ion{Na}{i} \citep{Wood2011, Zhou2012, Khalafinejad2018}, with transmission photometry detecting wavelength dependencies in $R_p$ (planetary radius) consistent with the potential presence of \ion{Na}{i} absorption \citep{Bento2014}. Using the FORS2 instrument at the VLT, \citet{Sedaghati2016} were unable to confirm the presence of \ion{Na}{i}, however, detected the wings of the \ion{K}{i} absorption feature (but not the line core, see Section \ref{section:oldstuff}), and ruled out a flat ``cloudy" spectrum. Analysis of WFC3/G141 observations by \citet{Mandell2013} detected strong H$_2$O absorption at 1.4\,\micron, and found that atmospheric models including a haze prescription produced a better fit to the spectrum than those without. However, it is important to note that the G141 data did not contain post transit information, and were conducted in stare mode as opposed to scan mode (see Sections 2.1 and 5.2 for further discussion).

WASP-17b was also analysed as part of the large comparative study by \citet{Sing2016} with new HST STIS measurements (GO-12473, PI Sing), \textit{Spitzer} IRAC data (90092, PI D\'esert), and a re-analysis of the WFC3/G141 measurements first analysed by \citet{Mandell2013} (GO-12181, PI Deming). \citet{Sing2016} concluded that of the 10 planets included in their study, WASP-17b had the clearest atmosphere, with prominent H$_2$O and \ion{Na}{i} absorption features (although no evidence of \ion{K}{i}), and IR transit depths higher than that of the optical, indicative of a lack of a strong optical slope. Through the strength of the \ion{Ca}{ii} H\&K emission lines as measured by Keck/HIRES, \citet{Sing2016} also found that the host star WASP-17 has a log$R' _{\mathrm{HK}}$ value of -5.531, indicative of a quiet star, a conclusion also found by \citet{Khalafinejad2018}.

WASP-17b has also been analysed extensively through theoretical and retrieval studies (\citealt{Barstow2017,Pinhas2019,Welbanks2019} using the transmission spectrum of \citealt{Sing2016}, and \citealt{Fisher2018} using the transmission spectrum of \citealt{Mandell2013}), with mixed evidence as to whether its transmission spectrum is best fit by cloudy or cloud-free models. \citet{Barstow2017} found that the atmosphere is best fit by a Rayleigh scattering aerosol, while \citet{Fisher2018} and \citet{Pinhas2019} found that both cloudy and cloud-free models are statistically comparable in fit. Reported H$_2$O abundances for WASP-17b have also spanned a wide range, with studies reporting sub-solar, solar and super-solar values \citep{Barstow2017, Fisher2018, Pinhas2019, Welbanks2019}. 

However, the WFC3/G141 observations on which many of these studies critically relied upon were undertaken before the spatial scanning technique became an available mode on HST. Performed in stare mode, the WFC3/G141 measurements, like all stare mode observations, were less efficient, more susceptible to systematics and achieved lower photometric precisions than can now be obtained with spatial scan observations. Furthermore, the WFC3/G141 observations additionally suffered due to a lack of complete transit coverage, failing to capture any egress or post-transit information. These measurements therefore do not provide as robust of constraints on the planet's spectrum as could be achieved with new observations. To that end, we present analysis of newly obtained WFC3/G102 and G141 data of WASP-17b taken in spatial scanning mode (GO-14918, PI Wakeford). We also present a comprehensive and consistent reanalysis of the STIS and \textit{Spitzer} observations, taking advantage of the many advances in analysis techniques which have occurred since the observations were featured in \citet{Sing2016}. These analyses allow us to present an extensive retrieval analysis of the panchromatic transmission spectrum of WASP-17b from 0.3--5\,$\micron$.

This paper is organised as follows. In Section \ref{section:reduction}, we present our observations and our data reduction process. In Section \ref{section:o-c}, we analyse TESS data and prior orbital ephemerides of WASP-17b, and present a new, refined orbital period. Section \ref{section:fitting} details the light curve fitting for each dataset. In Section \ref{section:results}, we present the results of the fitting, and our combined transmission spectrum is compared to previous works. In Section \ref{section:atmosphere}, we interpret the atmosphere of WASP-17b using results from both free and equilibrium chemistry atmospheric retrievals. Finally we present our conclusions in Section \ref{section:discuss_conc}.


\section{Observations \& Data Reduction} \label{section:reduction}

\subsection{HST/WFC3} \label{section:wfc3}

We observed two transits of WASP-17b with HST WFC3/IR as part of GO-14918 (PI Wakeford). HST observed one transit with the G102 grating (0.8 -- 1.1\,\micron) on UT 2017 June 16 (program visit 2), however as this visit suffered from guide star failure which rendered the data unusable, this observation was repeated on UT 2017 September 25 (program visit 32). A second transit was observed with the G141 grating (1.1 -- 1.6\,\micron) on UT 2017 July 23 (program visit 1). Both successful visits were conducted in forward spatial scan mode, with exposure times of 134.35\,s over five HST orbits, for a total of 70 exposures. Visits 1 and 32 achieved maximum pixel counts of $\sim39,000$ and $\sim32,000$ respectively. Both visits were read out using the SPARS25 sampling sequence with NSAMP=8.

To reduce our WFC3/IR observations, we used a custom pipeline built for this analysis, designed to work with the Exoplanet Timeseries Characterisation - Instrument Systematics Marginalisation light curve fitting package (\verb|ExoTiC-ISM|, \citealt{Laginja2020}). Our pipeline takes the dark and flat-field corrected \verb|ima| FITS products of the \verb|CALWF3| pipeline\footnote{http://www.stsci.edu/hst/wfc3/pipeline/} for each transit and performs cosmic ray removal, aperture selection, background removal and stellar spectra extraction. 

\begin{figure*}
    \centering
    \includegraphics[width=0.48\textwidth]{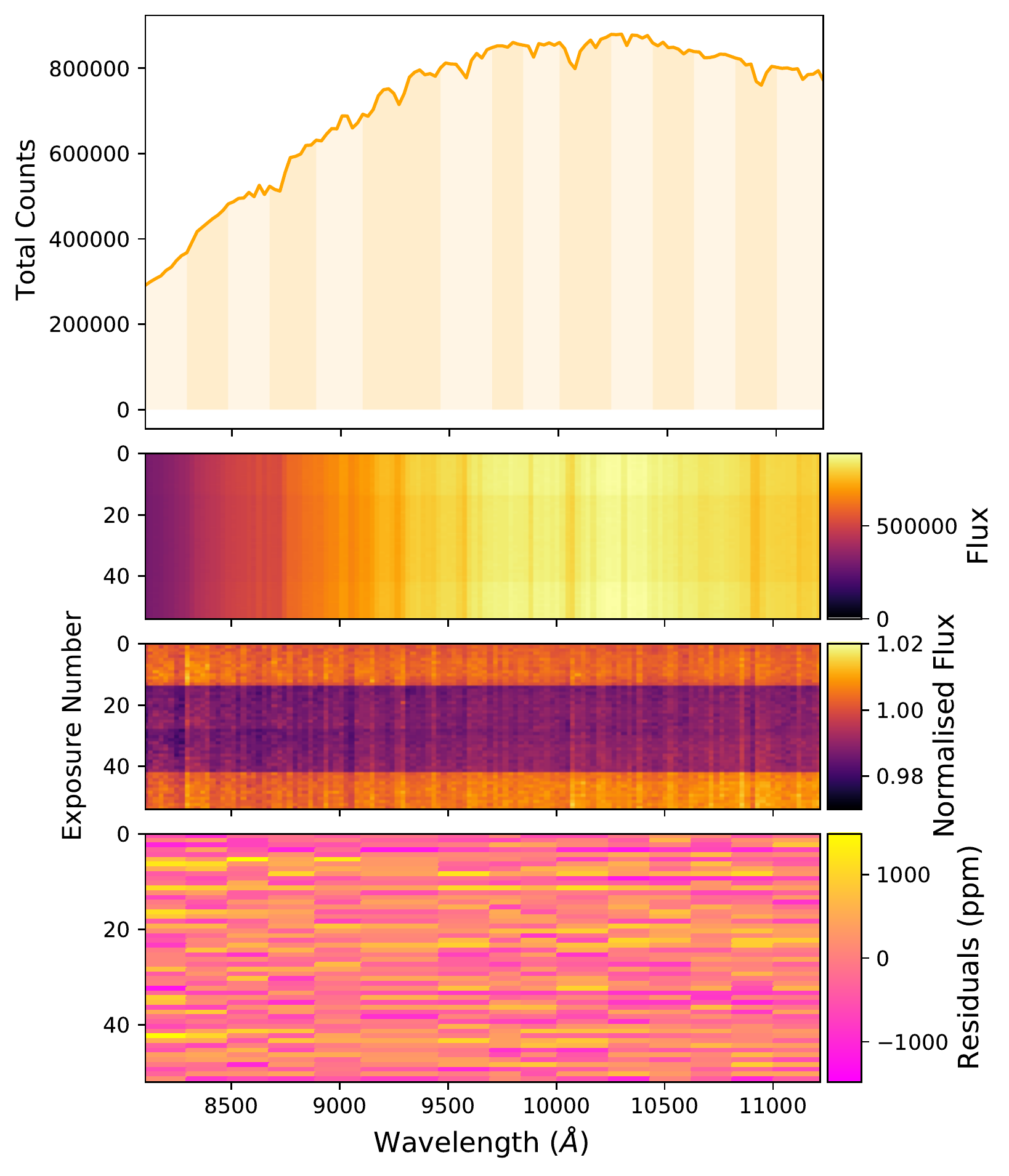}
    \includegraphics[width=0.45\textwidth]{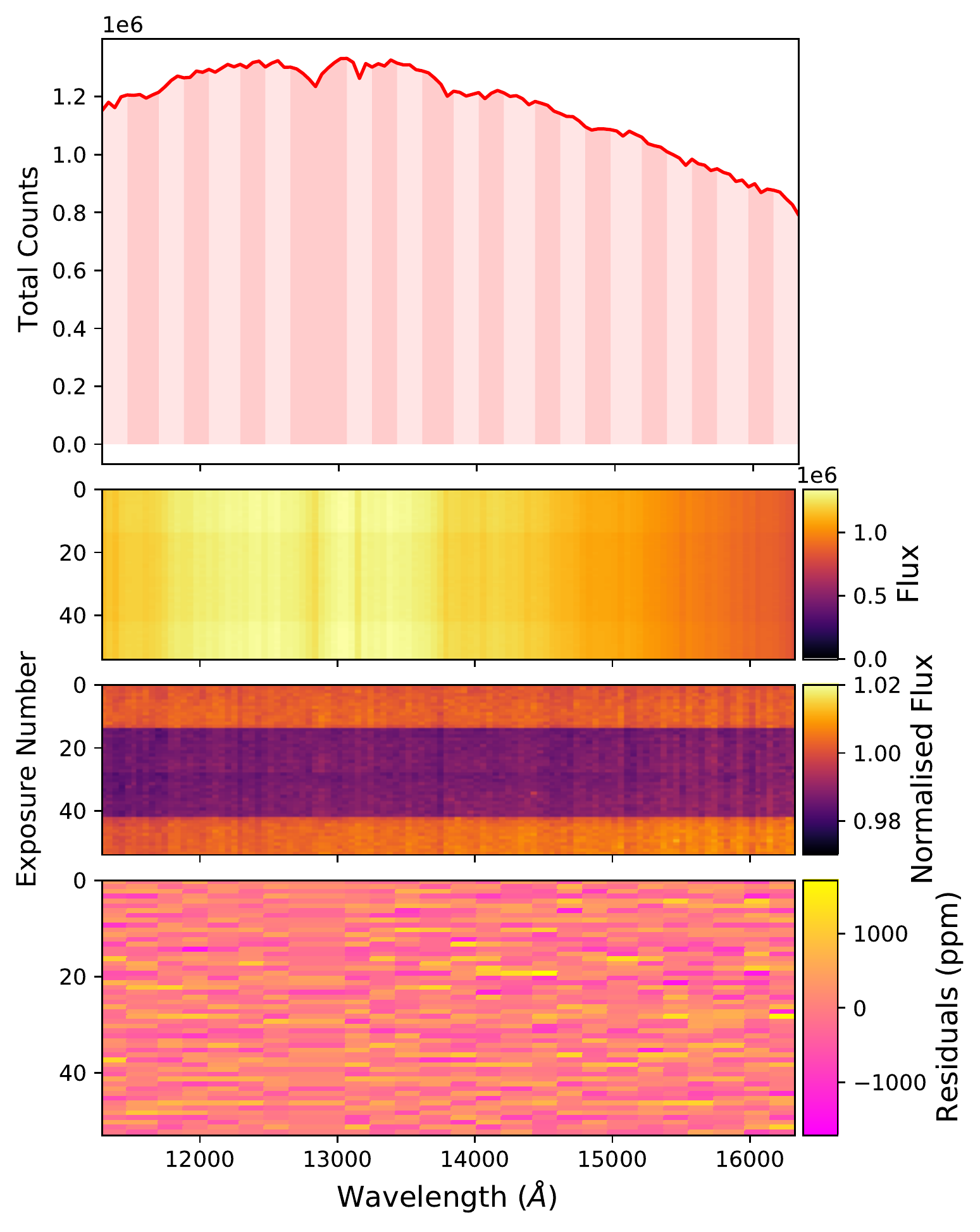}
    \caption{1D stellar spectrum and pixel maps for observations of WASP-17b from WFC3/G102 visit 32 (left, orange) and WFC3/G141 visit 1 (right, red). Top: Example stellar spectrum of WASP-17. Vertical bands show the wavelength bins corresponding to each spectroscopic light curve. Upper Middle: Stellar spectra from each visit following cosmic ray removal and extraction, but before systematic corrections, with units of total number of photons collected. Lower Middle: Same as above, but with stellar spectra normalised by the final spectrum to enhance the contrast of the transit. Bottom: Residuals of each spectroscopic light curve after systematic correction.}
    \label{fig:wfc3_maps}
\end{figure*}

We begin by flagging and removing any cosmic rays incident on the detector in the 2D images, replacing $>5\sigma$ outliers with the median value of that pixel across the time axis of the observations. This step is repeated iteratively a total of four times to ensure all cosmic rays are found and removed. We then extract the stellar spectra by summing the flux in each scan within an optimised aperture, found by minimising the standard deviations of post-transit white light curve data measured from a variety of aperture widths (see \citealt{Wakeford2016}). The determined aperture widths were 37 pixels wide for visit 1, and 30 pixels wide for visit 32. A background region is then selected with the same spatial width as the aperture, spanning the full image width in the dispersion direction, and the median count in this region is calculated and subtracted from each pixel. After extracting the spectra (shown in Figure \ref{fig:wfc3_maps}), we calibrate them in wavelength using target positioning offsets and detector calibration profiles \citep{g1412009,g1022009}, before calculating the $x$ sub-pixel shifts of the spectra over time by cross-correlation, for use in instrument systematic detrending \citep[e.g.,][]{Deming2013,Fraine2014,Sing2015}.

\begin{figure*}
    \centering
    \includegraphics[width=\linewidth]{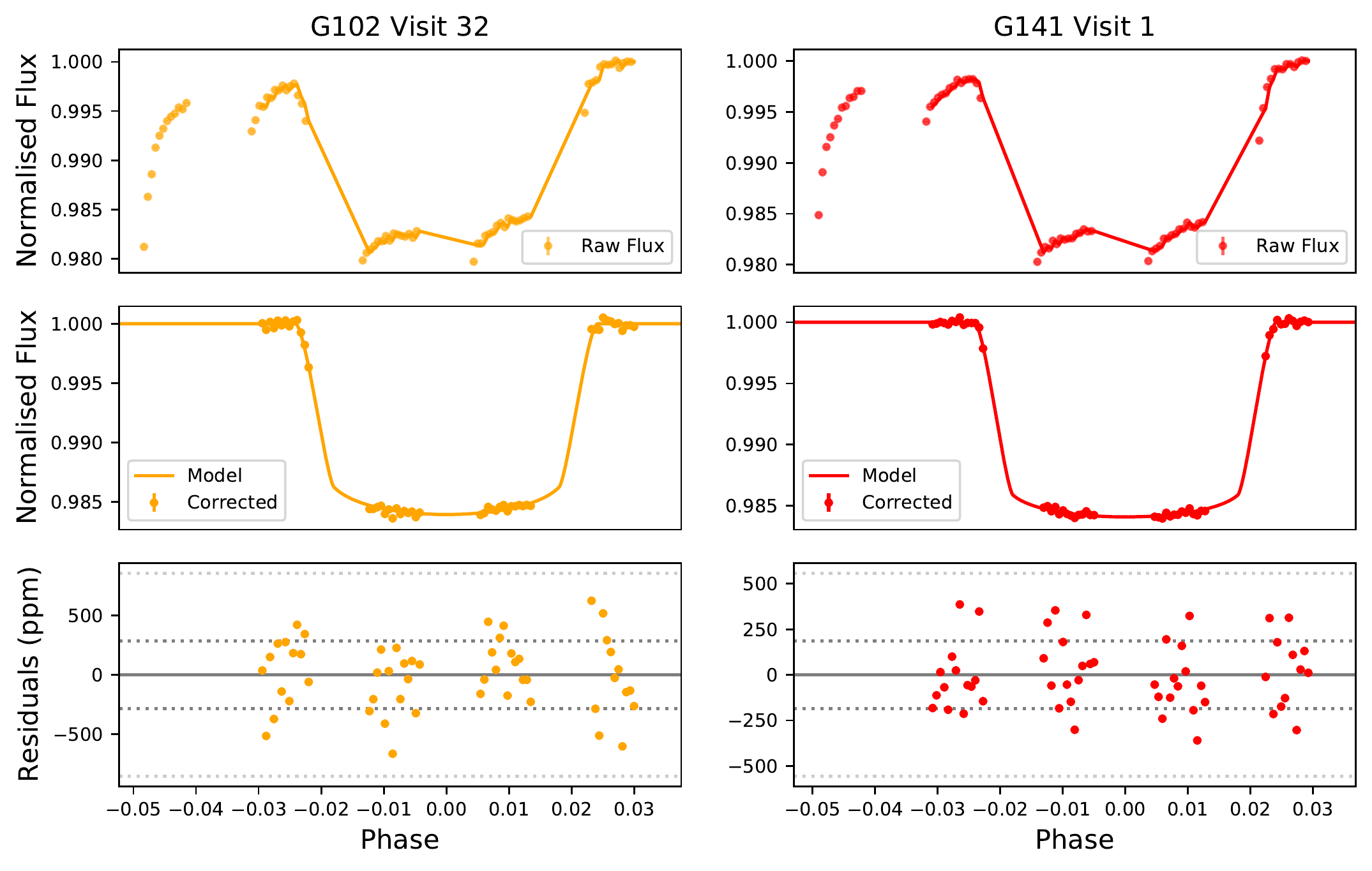}
    \caption{Transit white light curves of WASP-17b for the two HST WFC3 visits, G102 visit 32 (left) and G141 visit 1 (right). Top: The raw light curves. Data points from the zeroth orbit and first exposure in each orbit are plotted to demonstrate their differing systematic properties. Data points joined by lines are included in the light curve fitting. Middle: Corrected light curves and best fit model obtained with ExoTiC-ISM. Bottom: Corrected light curve residuals. The dashed lines indicate one and three times the standard deviation.}
    \label{fig:wfc3_wlcs}
\end{figure*}

We produce white light and spectroscopic light curves for each visit, discarding the zeroth orbit and first exposure in each orbit due to their significantly different systematics \citep[e.g.,][]{Deming2013, Wakeford2016, Zhou2017}. A band-integrated white light curve for each visit is produced by summing the flux of the stellar spectra across the full spectrum, as shown in Figure \ref{fig:wfc3_wlcs}. Spectroscopic light curves are produced by dividing the stellar spectra into wavelength bins with a minimum width of four pixels, equal to two resolution elements, and summing the fluxes of each bin. This reduces the effect of correlated noise between wavelength bins\footnote{https://hst-docs.stsci.edu/wfc3ihb/chapter-7-ir-imaging-with-wfc3/7-6-ir-optical-performance}. In total, 15 and 25 spectroscopic light curves were produced for the G102 and G141 observations respectively, resulting in spectral resolutions of $R\sim 90$ and $R\sim 140$ (see Figures \ref{fig:g102_slcs} \& \ref{fig:g141_slcs}).

\subsection{HST/STIS} \label{section:stis}

To present a consistent transmission spectrum with wide wavelength coverage, and to take advantage of advances in transit light curve analysis, we re-reduced and analysed three archival transit observations of WASP-17b from HST STIS as part of GO-12473 (PI D.K. Sing). Two transits were observed with the G430L grating (2892 -- 5700\,\AA) on UT 2012 June 08 (program visit 6) and 2013 March 18 (program visit 5), each with exposure times of 279\,s over five HST orbits and a total of 48 exposures and a maximum pixel count of $\sim$24,000. A single transit was observed with the G750L grating (5240 -- 10270\,\AA) on UT 2013 March 19 (program visit 19) with an exposure time of 259\,s over five HST orbits for a total of 48 exposures, with a maximum pixel count of $\sim$27,000. To minimise slit losses the 52$\times$2'' slit was used for both gratings. 

We reduced the STIS observations following the procedures outlined in \citet{Alam2018}. The raw 2D images for all STIS observations are corrected using the \verb|CALSTIS|\footnote{http://www.stsci.edu/hst/stis/software/analyzing/calibration/pipe\_soft\_
hist/intro.html} pipeline and relevant calibration files to account for bias, dark, and flat field corrections. We used a custom routine modelled on IRAF's APALL to extract the stellar spectrum from each exposure using the \verb|flt| science files. The G750L dataset was defringed using contemporaneous fringe flats \citep[see][for details]{Nikolov2014}. To identify and remove cosmic rays we followed the procedure outlined Section \ref{section:wfc3} with an additional spatial search in each frame. Due to the long exposure times and short wavelengths of the STIS observations, a large number of cosmic ray hits are expected to be measured. To determine the location of these, we took difference images for each frame, comparing these to the surrounding frames in time \citep{Nikolov2014}. We then took a median image of five frames and identified cosmic rays using a window of 20 pixels centred on each pixel, comparing the value of the central pixel to the median of all pixels in the window. Where the pixel value was $>5\sigma$ above the median level we flagged this as a cosmic ray and replaced the pixel value with a median of the column using the adjacent $\pm$\,3 pixels. 

\begin{figure*}
    \centering
    \includegraphics[width=0.33\linewidth]{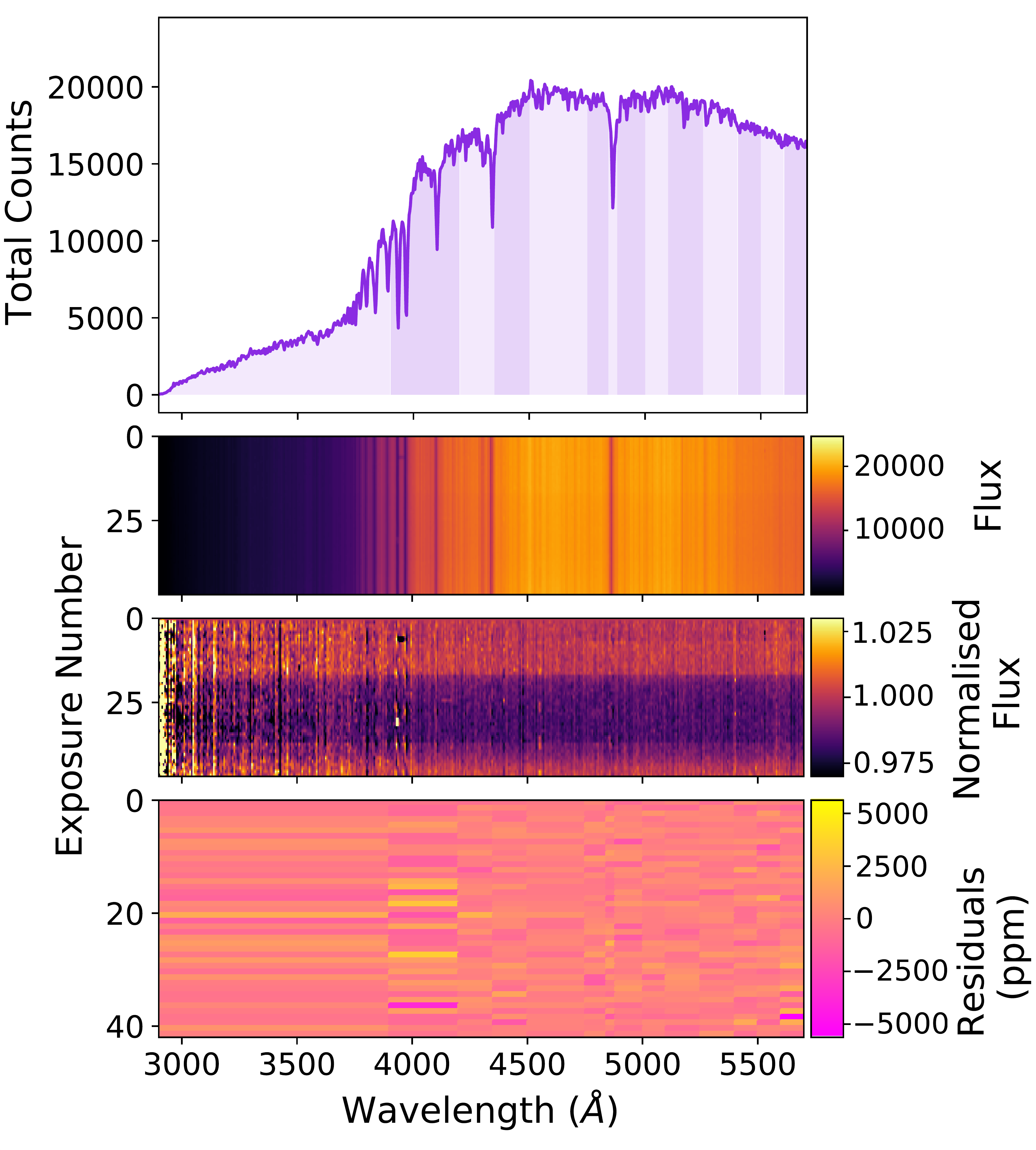}
    \includegraphics[width=0.33\linewidth]{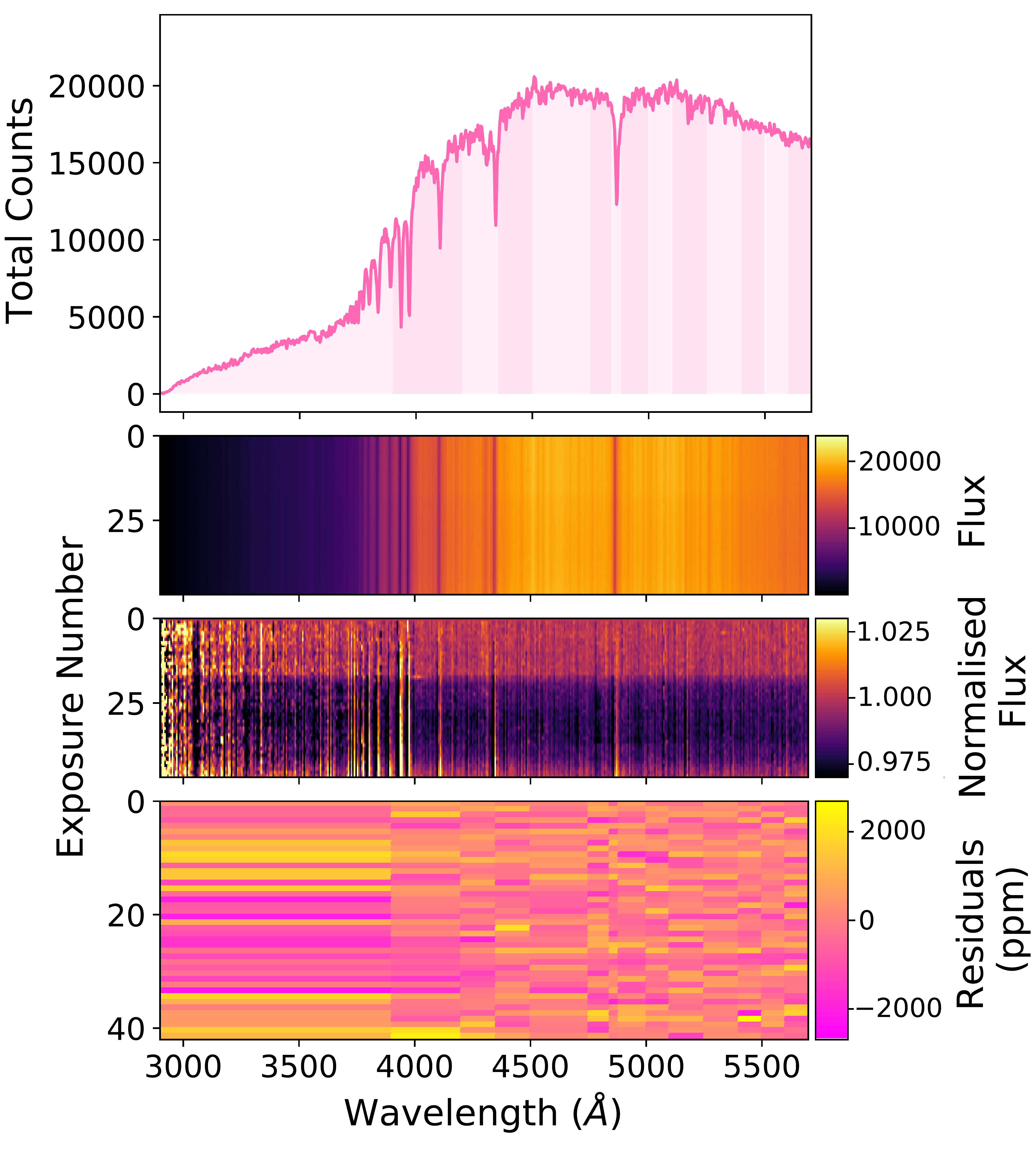}
    \includegraphics[width=0.33\linewidth]{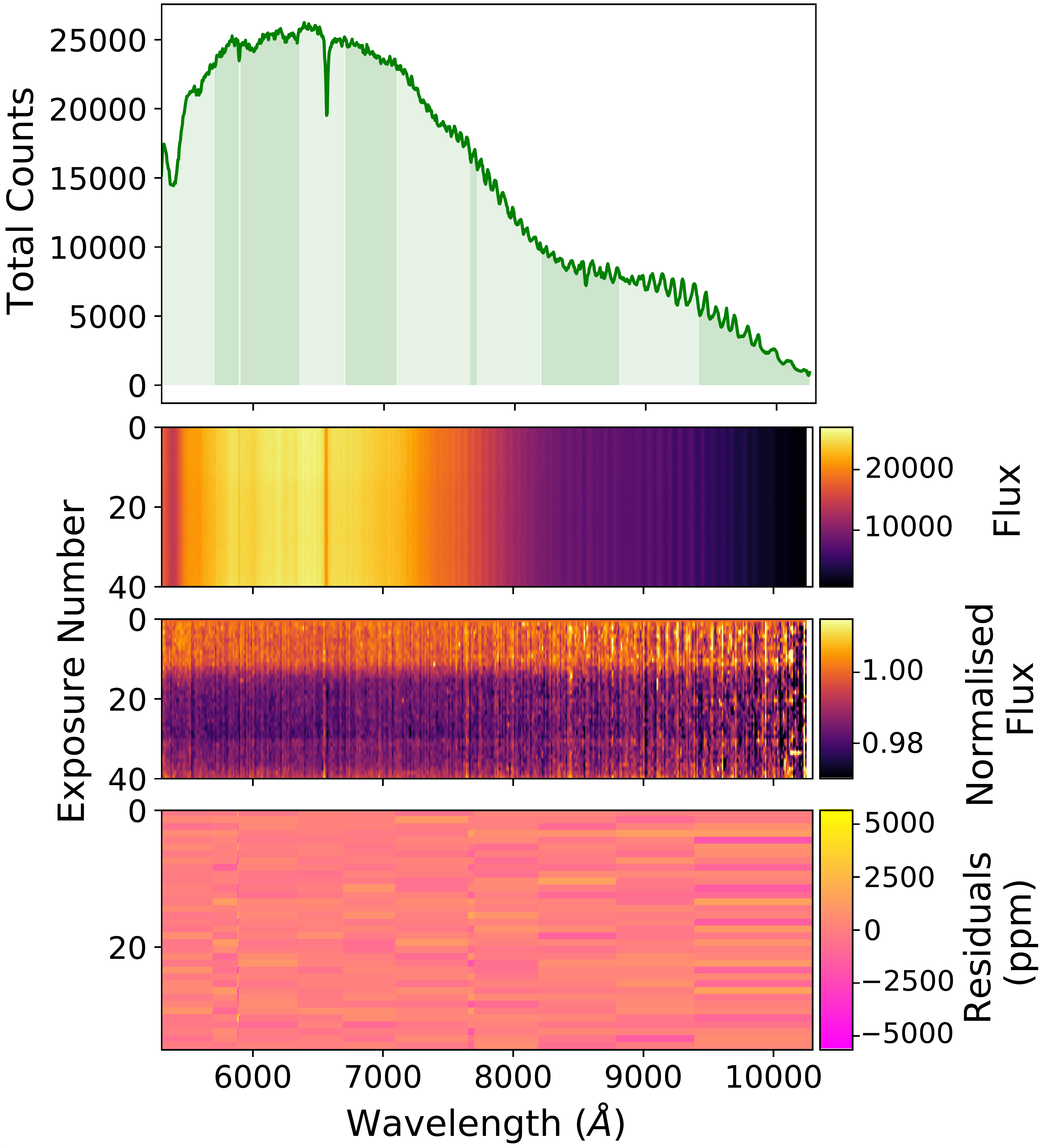}
    \caption{1D stellar spectrum and pixel maps for observations of WASP-17b from STIS/G430L visit 5 (left, purple) visit 6 (centre, pink) and STIS/G750L visit 19 (right, green). For details see Figure \ref{fig:wfc3_maps}.}
    \label{fig:stis_maps}
\end{figure*}

We tested a range of extraction aperture widths for each observation and found the lowest scatter with an aperture width of $\pm$\,6.5 pixels around the central trace on all three STIS observations. We defined the best aperture by minimising the standard deviation of the measured flux in orbit one of each observation. The wavelength solutions were computed by using the \verb|x1d| files from \verb|CALSTIS| to resample all of the extracted spectra (shown in Figure \ref{fig:stis_maps}) and cross correlate them to the common rest frame, taken as the final spectrum. From this we also obtain information on sub-pixel shifts in the dispersion direction due to the pointing motion of the telescope during the full transit observation. We use the measured shifts in this cross correlation in our systematic model to account for changes induced in the spectra over time.

\begin{figure*}
    \centering
    \includegraphics[width=\linewidth]{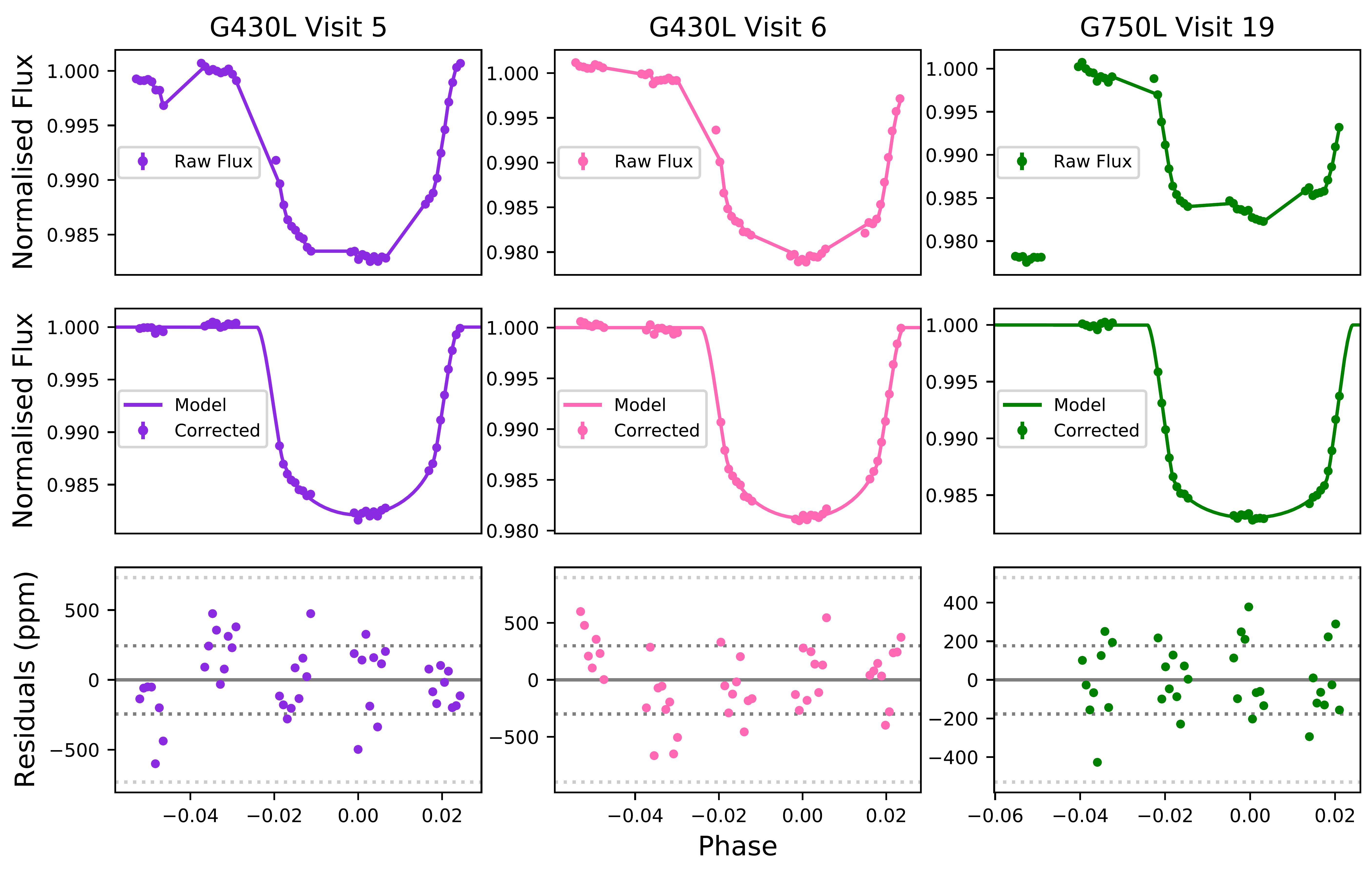}
    \caption{Transit white light curves of WASP-17b for the three HST STIS visits, G430L visit 5 (left), G430L visit 6 (middle) and G750L visit 19 (right). Top: The raw light curves. Data points corresponding to the first exposures in each orbit for all three visits, and from the zeroth orbit for G750L, are plotted to demonstrate their differing systematic properties. Data points joined by lines are included in the light curve fitting. Middle: Corrected light curves and best fit model obtained with a custom ExoTiC-ISM IDL routine. Bottom: Corrected light curve residuals. The dashed lines indicate one and three times the standard deviation.}
    \label{fig:stis_wlcs}
\end{figure*}

Band-integrated white light curves were produced for each visit by summing the flux over the whole wavelength range of the gratings, as shown in Figure \ref{fig:stis_wlcs} (2892 -- 5700\,\AA\ for G430L grating, 5240 -- 10270\,\AA\ for the G750L grating). Due to previously poor constraints on the planet's period and transit times, each of the STIS observations did not capture any post transit egress information on the stellar baseline (see Figure \ref{fig:stis_wlcs} and Section \ref{section:o-c} for details on the orbital period). 

As with the WFC3 observations, the first exposure in each orbit is discarded, as they are subject to different systematic effects to the subsequent exposures in that orbit. We choose to not discard the zeroth orbit of the two G430L observations as they do not display significant differences in their systematics to subsequent orbits. The inclusion of the zeroth orbits were found to improve the transit model fit to the data, likely due to the increases in the pre-transit baselines, as the observations are lacking in post-egress data. However, the zeroth orbit of the G750L observation did show significant differences in the flux compared to subsequent orbits and was therefore discarded for this analysis (see Figure \ref{fig:stis_wlcs}).

We create spectroscopic light curves using the wavelength bins published by \citet{Sing2016} (see Figures \ref{fig:g430_slcs} \& \ref{fig:g750_slcs}). These bins are chosen as they sample the \ion{Na}{i} and \ion{K}{i} lines covered by the G750L grating, while avoiding strong stellar lines and ensuring that the flux of the star is evenly sampled.

\subsection{\textit{Spitzer}/IRAC} \label{section:spitzer}

Continuing in our updated analysis, we re-reduced two archival transits of WASP-17b obtained by \textit{Spitzer}'s Infrared Array Camera \citep[IRAC,][]{Fazio2004} as a part of Program 90092 (PI: D\'esert), incorporating recent advances in exoplanet data reduction and light curve fitting techniques for \textit{Spitzer} observations. One transit was observed with the 3.6\,$\micron$ channel on UT 2013 May 10, and one transit was observed with the 4.5\,$\micron$ channel on UT 2013 May 14. Both observations had a frame time of 2 seconds, for a total of 14,720 frames in both channels. 

We used the Photometry for Orbits, Eclipses, and Transits  \citep[POET,][]{Campo2011,Stevenson2012, Cubillos2013} pipeline, including updates from \citet{May2020} which apply a fixed sensitivity map in order to remove the intrapixel sensitivity variations at 4.5\,$\micron$. The data is extracted using 2D Gaussian centroiding, with a fixed aperture size of 2.25 pixels for both transits. This aperture was determined by varying the aperture in 0.25 pixel increments between 2.00 and 4.00 pixels and selecting the size which results in the best signal difference to noise ratio (SDNR). We further use a fixed annulus between 7 and 15 pixels for background subtraction.

\section{Updated Orbital Period} \label{section:o-c}

\begin{figure}
    \centering
    \includegraphics[width=\linewidth]{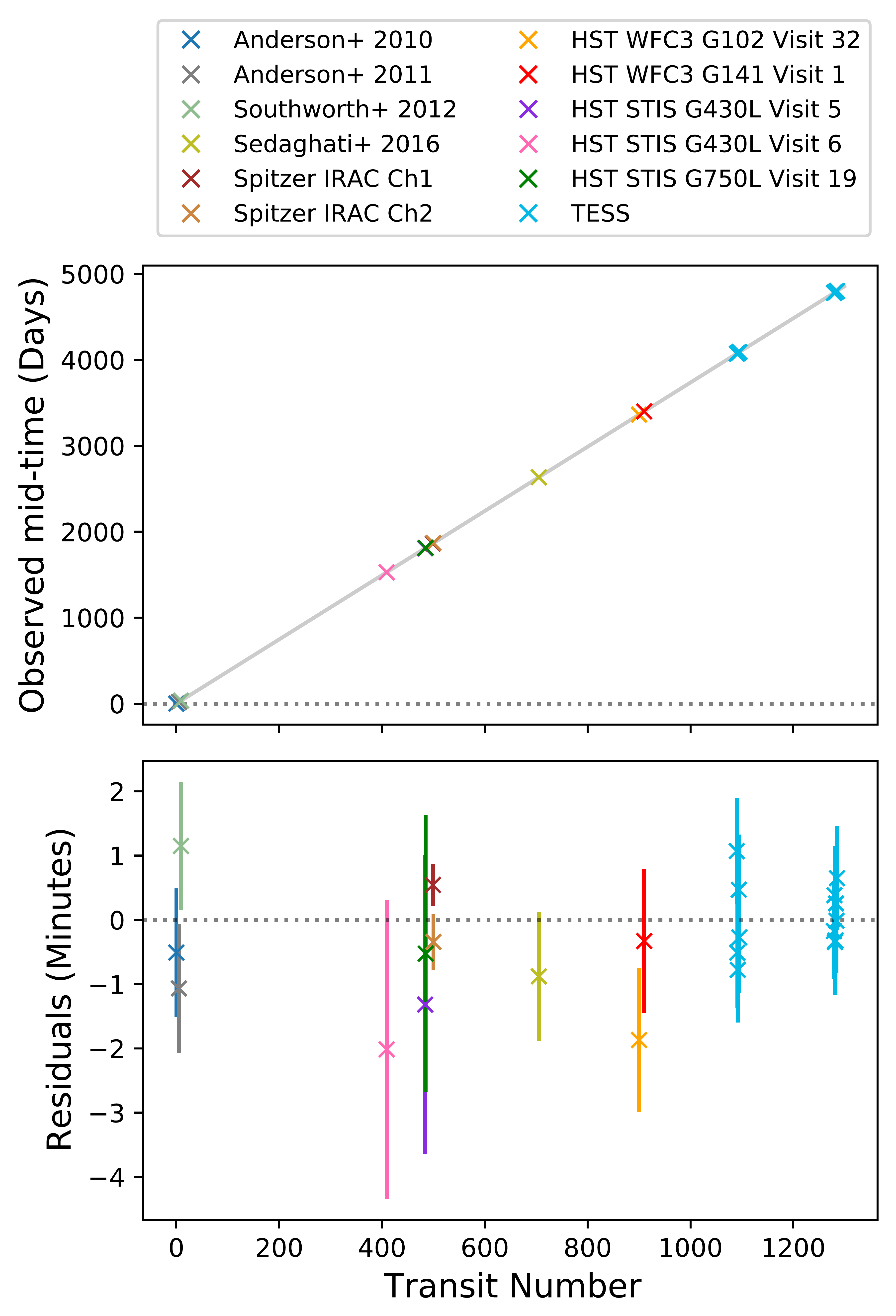}
    \includegraphics[width=\linewidth]{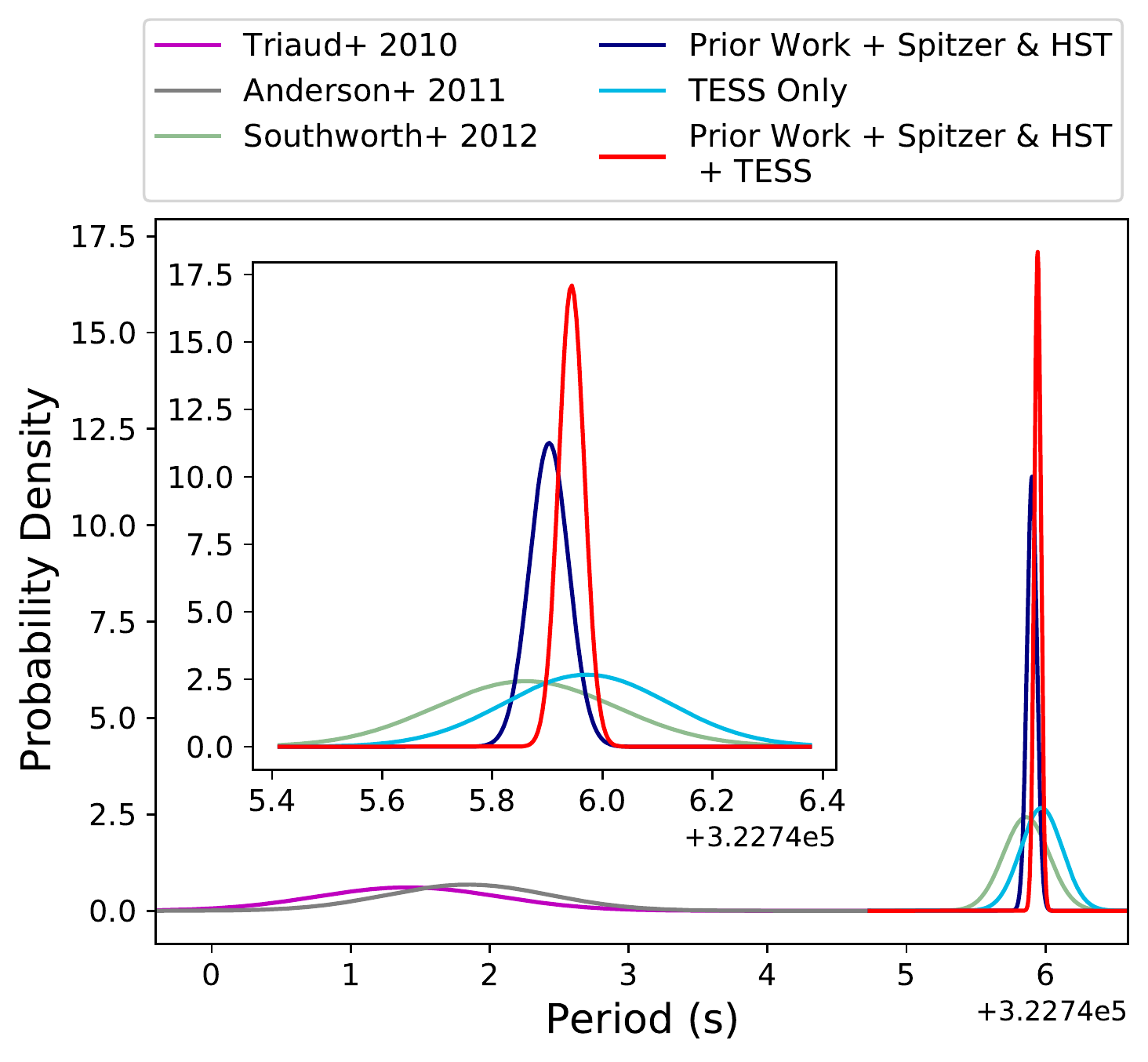}
    \caption{Top: Observed transit timings against calculated transit number, fitted with a linear relation. Middle: The residuals of this fit, giving the O-C diagram for previous mid-transit timings of WASP-17b. For details of the observations included, see Table \ref{tab:epoch_table}. Bottom: Distributions of the periods of \citet{Triaud2010}, \citet{Anderson2011} and \citet{Southworth2012}, along with the periods obtained when fitting different combinations of the values listed in Table \ref{tab:epoch_table}. All uncertainties are assumed to be Gaussian.}
    \label{fig:oc}
\end{figure}

\begin{table}
\centering
\caption{Mid-transit times used in Figure \ref{fig:oc} to calculate the period of WASP-17b. All times have been converted to BJD$_\mathrm{TDB}$.}
\label{tab:epoch_table}
\begin{tabular}{cccc}
\hline
Observation & Epoch (BJD$_\mathrm{TDB}$) (Days) \\
\hline
\citet{Anderson2010} & 2454559.18175 $\pm$ 0.000694 \\
\citet{Anderson2011} & 2454577.85879 $\pm$ 0.000694 \\
\citet{Southworth2012} & 2454592.802271 $\pm$ 0.000694 \\
\citet{Sedaghati2016} & 2457192.698749 $\pm$ 0.000694 \\
\textit{Spitzer} IRAC Ch1 & 2456423.18973 $\pm$ 0.00023 \\
\textit{Spitzer} IRAC Ch2 & 2456426.9246 $\pm$ 0.0003 \\
HST WFC3 G102 Visit 32 & 2457921.11772783 $\pm$ 0.000775 \\
HST WFC3 G141 Visit 1 & 2457958.473652 $\pm$ 0.000775 \\
HST STIS G430L Visit 5 & 2456367.15615529 $\pm$ 0.001615 \\
HST STIS G430L Visit 6 & 2456086.99426107 $\pm$ 0.001615 \\
HST STIS G750L Visit 19 & 2456370.8921914 $\pm$ 0.001499 \\
TESS Sector 12 Transit 1 & 2458630.86200815 $\pm$ 0.000575 \\
TESS Sector 12 Transit 2 & 2458634.59639647 $\pm$ 0.000596 \\
TESS Sector 12 Transit 3 & 2458638.33169591 $\pm$ 0.000569 \\
TESS Sector 12 Transit 4 & 2458645.80353243 $\pm$ 0.000597 \\
TESS Sector 12 Transit 5 & 2458649.53850306 $\pm$ 0.000596 \\
TESS Sector 38 Transit 1 & 2459336.86789830 $\pm$ 0.000514 \\
TESS Sector 38 Transit 2 & 2459340.60377057 $\pm$ 0.000530 \\
TESS Sector 38 Transit 3 & 2459344.33875063 $\pm$ 0.000571 \\
TESS Sector 38 Transit 4 & 2459348.07425237 $\pm$ 0.000592 \\
TESS Sector 38 Transit 5 & 2459351.81013867 $\pm$ 0.000548 \\
TESS Sector 38 Transit 6 & 2459355.54543725 $\pm$ 0.000561 \\
TESS Sector 38 Transit 7 & 2459359.28138246 $\pm$ 0.000563 \\
\hline
\hline
Fitted Period (Days): & 3.73548546 $\pm$ 0.00000027 \\
\hline
\end{tabular}
\end{table}

In preparation for light curve fitting, here we revisit and update WASP-17b's orbital period. After assessing the existing literature values for the system parameters, we found that there is a discrepancy in published periods of WASP-17b, with reported periods separated into two distinct groups $\sim$4\,s different from one another. As illustrated in Figure \ref{fig:oc}, several studies report periods within $1\sigma$ of 3.735435\,d \citep[e.g.,][]{Anderson2010, Anderson2011, Sedaghati2016}, while other studies \citep[e.g.,][]{Southworth2012} and initial fitting of the HST data obtain periods $\sim$4\,s longer, within $1\sigma$ of 3.735485\,d. While this difference is only 0.001\% of the period of WASP-17b, accurate transit timings are a key element in observation planning and light curve fitting, particularly in the case of HST where there are significant gaps in observations during single transit events. Poor constraints on a planet's period and transit timings can lead to observations which miss pre- or post-transit information (e.g., impacting the STIS observations as outlined in Section \ref{section:stis}), an effect which will continue to worsen if the orbital ephemerides are not updated.

\subsection{TESS} \label{section:tess}

To obtain a longer temporal baseline for our refinement of the orbital period of WASP-17b, we made use of transits observed by the Transiting Exoplanet Survey Satellite (TESS \citealt{Ricker2014}).

During its primary mission, TESS observed 5 transits of WASP-17b (TIC 66818296) in sector 12, with a further 7 transits observed during the extended mission in sector 38. The 12 total transits observed by TESS substantially increase the baseline available for transit timing measurements (see Section \ref{section:diagram}). We downloaded the WASP-17b data from the MAST data archive\footnote{https://archive.stsci.edu/} using the python package \verb|lightkurve| \citep{Lightkurve2018}, before stitching the two sectors together, normalising the light curve by its median value, and selecting the PDCSAP flux \citep{Smith2012} as the starting point of our analysis. All the data has a cadence of 2 minutes, and we convert the times from BJTD (Barycentric TESS Julian Date) to $\rm{BJD}_{\rm{TDB}}$.

We initially fit the entire light curve to generate a period estimate, found to be 3.73548578\,d, before fixing this value and individually fitting 1-day windows of data centred on each transit, with the aim of measuring the mid-transit times. For all of our TESS lightcurve fits, we use the python package \verb|batman| \citep{Kreidberg2015} to model the transits, and the fast 1D Gaussian Process (GP) code, \verb|celerite| \citep{Foreman-Mackey2017, Foreman-Mackey2018}, to model any systematics. In the transit model, we allow the mid-transit time ($T_0$), ratio of the planet to stellar radii ($R_p/R_s$), ratio of the semi-major axis to stellar radius ($a/R_s$), and inclination ($i$) to vary, and set the eccentricity to zero \citep{Southworth2012}. We employ a four-parameter non-linear limb-darkening law, with the coefficients fixed to values computed using the Exoplanet Characterisation Toolkit (ExoCTK, \citealt{exoctk2021}). For the systematics model, we use an approximate Matern-3/2 kernel, parameterised by a length scale and an amplitude, both of which are marginalised over during the fitting. We include one additional free parameter, a constant variance term, which is added to the diagonal of the covariance matrix to account for underestimated uncertainties, however we find this parameter is constrained to negligible values. Further tests were performed to assess the sensitivity of the results to the transit model parameterisation, the size of the data window around each transit, testing 1 and 2 day windows, and GP kernel type, testing a stochastically driven, damped, harmonic oscillator kernel. We also test various limb-darkening laws, with both fixed and free parameters. In all these tests we find the results are insensitive to these choices.

The fitting is performed using the affine-invariant Markov Chain Monte Carlo (MCMC) algorithm, \verb|emcee| \citep{Foreman-Mackey2013}. The resulting mid-transit times from all 12 TESS transits are shown in Table \ref{tab:epoch_table}, where the uncertainties are the $68$\% credible intervals of the period posterior distributions. These mid-transit times are later used to refine the orbital period of WASP-17b, as described in Section \ref{section:diagram}. The final fit to all the TESS data using our updated orbital period is shown in Figure \ref{fig:tess_folded_lc}.

\subsection{Refining WASP-17b's Orbital Period}
\label{section:diagram}

We used the mid-transit times of WASP-17b from previous observations and those presented in this work (see Table \ref{tab:epoch_table}) to produce an observed - calculated mid-transit timing (O-C) diagram in order to better refine the planet's orbital period. We converted all available transit times to BJD$_\mathrm{TDB}$ using the tools and methods outlined by \citet{Eastman2010}, and fitted a linear relation to the plot of observed transit times against the transit number \citep{Agol2018}, to obtain an updated period, as shown in Figure \ref{fig:oc}. The resulting period is 3.73548546 $\pm$ 0.00000027\,d, and we henceforth adopt this value for this work.

To visualise the improvement in the orbital period, Figure \ref{fig:oc} shows the probability densities of a selection of periods from previous works \citep{Triaud2010, Anderson2011, Southworth2012}, along with the periods obtained by fitting different combinations of the transit timings of the data presented in Table \ref{tab:epoch_table}. Assuming Gaussian uncertainties, we plot normal distributions centred on the periods with standard deviations of the quoted uncertainties. Due to the significant increase in transit timings now available, we are able to improve upon previous constraints, and thanks to the extended baseline provided by TESS, see the best refinement of the period uncertainty when considering all available data.

\section{Light Curve Fitting} \label{section:fitting}

To extract the 0.3 -- 5.0\,\micron\, transmission spectrum of WASP-17b, we fit our light curves to obtain their respective transit depths and relative uncertainties. Our fitting procedure for the HST STIS and WFC3 white and spectroscopic light curves is described in Section \ref{section:hst_fitting}. Our fitting procedure, along with Bilinearly Interpolated Subpixel Sensitivity (BLISS) mapping and PRF detrending for the \textit{Spitzer} IRAC photometric light curves is described in Section \ref{section:spitzer_fitting}.

\begin{table}
\centering
\caption{Star and Planet parameters held fixed in the light curve fitting.}
\label{tab:lc_params}
\begin{tabular}{ccc}
\hline
Parameter & Value & Reference \\ 
\hline
[Fe/H] (dex) & -0.25 & \citet{Southworth2012} \\
$T_\mathrm{eff}$ (K) & 6550 & \citet{Southworth2012} \\ 
log(g) & 4.2 & \citet{Southworth2012} \\
$a/R_*$ & 7.025 & \citet{Sedaghati2016} \\
Eccentricity & 0 & \citet{Sedaghati2016} \\
& & Weighted Mean of \\
Inclination ($^\circ$) & 86.9 &  \citet{Anderson2010, Anderson2011}  \\
& & \& \citet{Sedaghati2016} \\
Period (Days) & 3.73548546 & This Work \\
\hline
\end{tabular}
\end{table}

\subsection{\textit{Hubble}}\label{section:hst_fitting}

To fit the HST light curves and correct them for systematic effects from the telescope and instruments, we use the systematic instrument marginalisation method outlined in \citet{Wakeford2016}. For our WFC3/G102 and G141 light curves, we use the \verb|ExoTiC-ISM| python package developed by \citet{Laginja2020}. \verb|ExoTiC-ISM| uses a Levenberg-Marquardt least-squares minimisation over a grid of 50 systematic models\footnote{https://github.com/Exo-TiC/ExoTiC-ISM\#the-systematic-model-grid} to obtain a set of fitted transit parameters for each model, making use of the resulting Akaike Information Criterion (AIC, \citealt{Akike1974}) to calculate each model's evidence and weight. The weights can then be used to calculate marginalised fit parameters, resulting in robust transit depths that do not heavily depend on the choice of an individual systematic model, and, as shown in previous work, obtains spectra consistent with other approaches \citep[e.g.,][]{Wakeford2016, Wakeford2020}. 

As \verb|ExoTiC-ISM| is currently only set up for WFC3 systematics, to analyse the STIS light curves we used a custom IDL (Interactive Data Language, \citealt{IDL}) routine to implement the systematic models used in \citet{Wakeford2017Sci}. These models account for a linear slope in time ($\theta$), HST thermal breathing ($\phi$), shifts in wavelength position ($\delta_\lambda$), and the x and y positional shifts on the detector throughout the observation, required due to the use of the slit in STIS spectra. The STIS systematic models take the form
\begin{equation}
    S(t, \lambda) = T_1 \theta + \sum_{i=1}^{4}p_i \phi^i + l_1 \delta_\lambda + a_1 x + a_2 y \mathrm{,}
\end{equation}
where $T_1$, $p_i$, $l_1$, $a_1$, and $a_2$ are coefficients fixed to zero or free parameters in the systematic model being fit. In total we fit 80 systematic models for each STIS light curve, computing the evidence for each model based on the AIC. The evidence is then used to compute a weighting for each model which is used to marginalise the results to obtain the marginalised transit depth and uncertainty following the steps outlined in \citet{Wakeford2016} and processed using the \verb|ExoTiC-ISM| framework. 

In both the WFC3 and STIS light curves we account for stellar limb darkening using a 4-parameter non-linear limb-darkening law \citep{claret2000,sing2010} and, due to the lack of phase coverage, fix the values to those derived from the 3D stellar models presented in \citet{magic2015}, using the stellar parameters shown in Table\,\ref{tab:lc_params}. The stellar models can be found in the \verb|ExoTiC-ISM| python package where they can be implemented for both WFC3 and STIS spectroscopic instruments.

\subsubsection{White Light Curves} \label{section:hst_wlcs}

For both WFC3 and STIS, we first fit the white light curves, fitting for the mid-transit time, transit depth and baseline stellar flux, and holding $a/R_{*}$, eccentricity and inclination fixed to literature values, with the period fixed to the new value (see Section \ref{section:o-c}, as outlined in Table \ref{tab:lc_params}). For a more accurate fit, we computed the weighted mean of the inclinations presented in \citet{Anderson2010, Anderson2011} and \citet{Sedaghati2016}, and hold the inclination fixed to this value. The raw and resulting best fit WFC3 and STIS white light curves are shown in Figures \ref{fig:wfc3_wlcs} and \ref{fig:stis_wlcs}, while the resulting transit depths are given in Table \ref{tab:wlc_depths}.

\begin{table}
\centering
\caption{Results of the white light curve fits for all visits and relative uncertainties.}
\label{tab:wlc_depths}
\begin{tabular}{cccc}
\hline

Instrument & Visit & Transit & Error \\
 & Number &  Depth (\%) & (\%) \\
\hline
HST WFC3 G102 & 32 & 1.4778 & 0.0028 \\
HST WFC3 G141 & 1 & 1.4899 & 0.0024 \\
HST STIS G430L & 5 & 1.5081 & 0.0150 \\
HST STIS G430L & 6 & 1.5708 & 0.0161 \\
HST STIS G750L & 19 & 1.5023 & 0.0276 \\
\textit{Spitzer} IRAC & Ch1 & 1.5177 & 0.0123 \\
\textit{Spitzer} IRAC & Ch2 & 1.5679 & 0.0149 \\
TESS Combined & - & 1.5103 & 0.0121 \\
\hline
\end{tabular}
\end{table}

\subsubsection{Spectroscopic Light Curves} \label{section:hst_slcs}

We next fit the spectroscopic light curves, following the same procedures as for the white light curve, marginalising over the systematic grids for each spectroscopic light curve individually. The residuals of these fits are shown in the lower panels of Figures \ref{fig:wfc3_maps} and \ref{fig:stis_maps} for our WFC3 and STIS observations respectively. The middle panels of Figures \ref{fig:wfc3_maps} and \ref{fig:stis_maps} also show the pixel maps of the stellar spectra once the extraction process is complete. These plots demonstrate our high quality low noise data and extraction and fitting routines, with no obvious bad pixels in the stellar spectra, and no wavelength dependent trends in the spectroscopic light curves.

The measured transit depth values and errors for each spectroscopic channel are shown in Tables \ref{tab:wfc3_table}, \ref{tab:g750_table} and \ref{tab:g430_table}. Our final transmission spectrum combining the information from all five visits can be seen in Figure \ref{fig:transmission}, with the combined transit depth and error of the two G430L visits given by their weighted mean, as shown in Table \ref{tab:g430_table}.

\begin{table}
\centering
\caption{Results of the wavelength-binned fits for WFC3/G102 and G141 visits (Figures \ref{fig:g102_slcs} \& \ref{fig:g141_slcs}) and relative uncertainties.}
\label{tab:wfc3_table}
\begin{tabular}{cccc}
\hline
$\lambda$ & $\Delta \lambda$ & Transit & Error \\
(\micron) & (\micron) &  Depth (\%) & (ppm) \\
\hline
 \multicolumn{4}{c}{-- G102 --}\\
0.81979 & 0.01902 & 1.50766 & 0.0162 \\
0.83881 & 0.01902 & 1.48501 & 0.0139 \\
0.85783 & 0.01902 & 1.49152 & 0.0132 \\
0.87803 & 0.02140 & 1.46812 & 0.0115 \\
0.89943 & 0.02140 & 1.47618 & 0.0109 \\
0.92796 & 0.03566 & 1.46501 & 0.0080 \\
0.95768 & 0.02377 & 1.48400 & 0.0093 \\
0.97670 & 0.01426 & 1.48733 & 0.0115 \\
0.99215 & 0.01664 & 1.46927 & 0.0107 \\
1.01236 & 0.02377 & 1.47548 & 0.0090 \\
1.03376 & 0.01902 & 1.46933 & 0.0099 \\
1.05278 & 0.01902 & 1.44943 & 0.0101 \\
1.07180 & 0.01902 & 1.47691 & 0.0104 \\
1.09087 & 0.01902 & 1.50017 & 0.0105 \\
1.11103 & 0.02140 & 1.50330 & 0.0098 \\
 \multicolumn{4}{c}{-- G141 --}\\
1.13801 & 0.01816 & 1.51271 & 0.0118 \\
1.15845 & 0.02270 & 1.49809 & 0.0111 \\
1.17888 & 0.01816 & 1.47967 & 0.0113 \\
1.19704 & 0.01816 & 1.47529 & 0.0117 \\
1.21747 & 0.02270 & 1.48210 & 0.0104 \\
1.23790 & 0.01816 & 1.47744 & 0.0109 \\
1.25606 & 0.01816 & 1.45743 & 0.0116 \\
1.28557 & 0.04086 & 1.47289 & 0.0081 \\
1.31509 & 0.01816 & 1.49151 & 0.0114 \\
1.33325 & 0.01816 & 1.49072 & 0.0114 \\
1.35141 & 0.01816 & 1.51744 & 0.0123 \\
1.37184 & 0.02270 & 1.50988 & 0.0112 \\
1.39227 & 0.01816 & 1.52036 & 0.0119 \\
1.41043 & 0.01816 & 1.52419 & 0.0118 \\
1.43086 & 0.02270 & 1.51715 & 0.0109 \\
1.45130 & 0.01816 & 1.51481 & 0.0121 \\
1.46946 & 0.01816 & 1.53627 & 0.0129 \\
1.48762 & 0.01816 & 1.52800 & 0.0131 \\
1.50805 & 0.02270 & 1.51300 & 0.0117 \\ 
1.52848 & 0.01816 & 1.50233 & 0.0137 \\
1.54664 & 0.01816 & 1.49508 & 0.0132 \\
1.56480 & 0.01816 & 1.45936 & 0.0133 \\
1.58524 & 0.02270 & 1.46988 & 0.0135 \\
1.60567 & 0.01816 & 1.47278 & 0.0145 \\
1.62383 & 0.01816 & 1.44110 & 0.0148 \\
\hline
\end{tabular}
\end{table}

\begin{table}
\centering
\caption{Results of the wavelength-binned fits for STIS/G750L visit 19 (Figure \ref{fig:g750_slcs}) and relative uncertainties.}
\label{tab:g750_table}
\begin{tabular}{cccc}
\hline
$\lambda$ & $\Delta \lambda$ & Transit & Error \\
(\micron) & (\micron) &  Depth (\%) & (ppm) \\
\hline
0.55000 & 0.04000 & 1.44694 & 0.0629 \\
0.57940 & 0.01880 & 1.48929 & 0.0621 \\
0.58930 & 0.00100 & 1.55754 & 0.1999 \\
0.61240 & 0.04520 & 1.52691 & 0.0422 \\
0.65250 & 0.03500 & 1.46307 & 0.0448 \\
0.69000 & 0.04000 & 1.51113 & 0.0429 \\
0.73800 & 0.05600 & 1.50697 & 0.0401 \\
0.76840 & 0.00480 & 1.50241 & 0.1365 \\
0.79540 & 0.04920 & 1.54974 & 0.0789 \\
0.85000 & 0.06000 & 1.47629 & 0.0660 \\
0.91000 & 0.06000 & 1.43743 & 0.0692 \\
0.98500 & 0.09000 & 1.47111 & 0.0853 \\
\hline
\end{tabular}
\end{table}

\begin{table*}
\centering
\caption{Results of the wavelength-binned fits for STIS/G430L visits 5 and 6 (Figure \ref{fig:g430_slcs}) and the weighted average transit depths, along with their relative uncertainties.}
\label{tab:g430_table}
\begin{tabular}{cccc|cc|cc}
\hline
$\lambda$ & $\Delta \lambda$ & Visit 5 & Error & Visit 6 & Error & \textbf{Combined} & \textbf{Error} \\
(\micron) & (\micron) & Transit Depth (\%) & (ppm) & Transit Depth (\%) & (ppm) & \textbf{Transit Depth (\%)} & \textbf{(ppm)} \\
\hline
0.34000 & 0.10000 & 1.55130 & 386 & 1.61964 & 419 & 1.58547 & 285 \\
0.40500 & 0.03000 & 1.46790 & 385 & 1.63700 & 363 & 1.55245 & 265 \\
0.42750 & 0.01500 & 1.55149 & 400 & 1.62705 & 412 & 1.58927 & 288 \\
0.44250 & 0.01500 & 1.49618 & 348 & 1.58308 & 391 & 1.53963 & 262 \\
0.46250 & 0.02500 & 1.48968 & 272 & 1.54781 & 253 & 1.51874 & 186 \\
0.47960 & 0.00920 & 1.50913 & 471 & 1.63180 & 498 & 1.57046 & 343 \\
0.48610 & 0.00400 & 1.55433 & 716 & 1.57314 & 691 & 1.56374 & 498 \\
0.49410 & 0.01200 & 1.49842 & 372 & 1.51725 & 400 & 1.50784 & 273 \\
0.50505 & 0.00990 & 1.54108 & 425 & 1.55753 & 482 & 1.54931 & 321 \\
0.51750 & 0.01500 & 1.49238 & 345 & 1.49490 & 347 & 1.49364 & 244 \\
0.53250 & 0.01500 & 1.52485 & 370 & 1.52233 & 349 & 1.52359 & 255 \\
0.54500 & 0.01000 & 1.49947 & 430 & 1.53033 & 434 & 1.51490 & 305 \\
0.55500 & 0.01000 & 1.53247 & 465 & 1.52748 & 484 & 1.52998 & 335 \\
0.56500 & 0.01000 & 1.58632 & 665 & 1.47848 & 623 & 1.53240 & 456 \\
\hline
\end{tabular}
\end{table*}

\subsection{\textit{Spitzer} Photometry} \label{section:spitzer_fitting}

\begin{figure}
    \centering
    \includegraphics[width=0.75\linewidth,trim={0.5cm 0.5cm 0.7cm 1cm},clip]{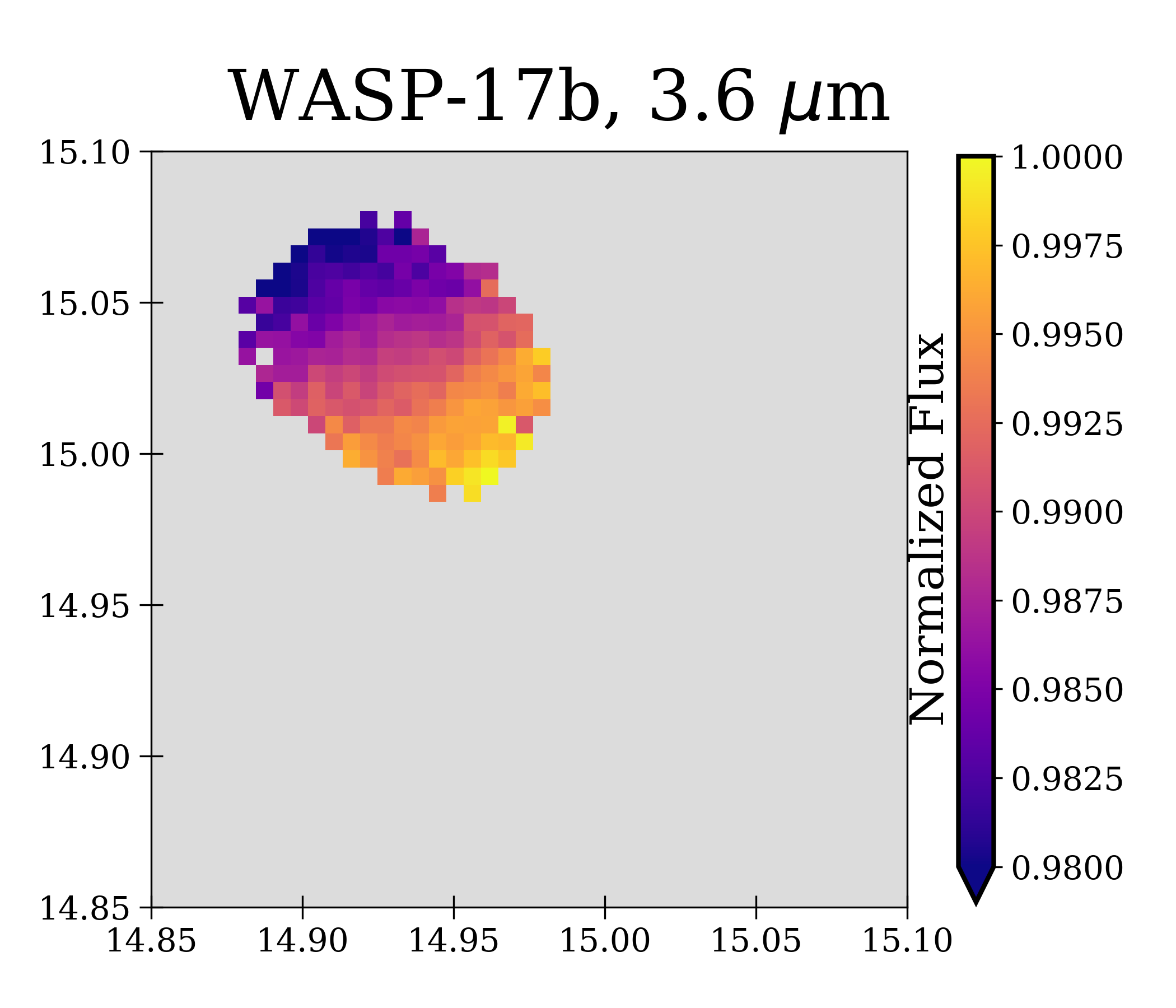}
    \includegraphics[width=0.75\linewidth,trim={0.5cm 0.5cm 0.7cm 1cm},clip]{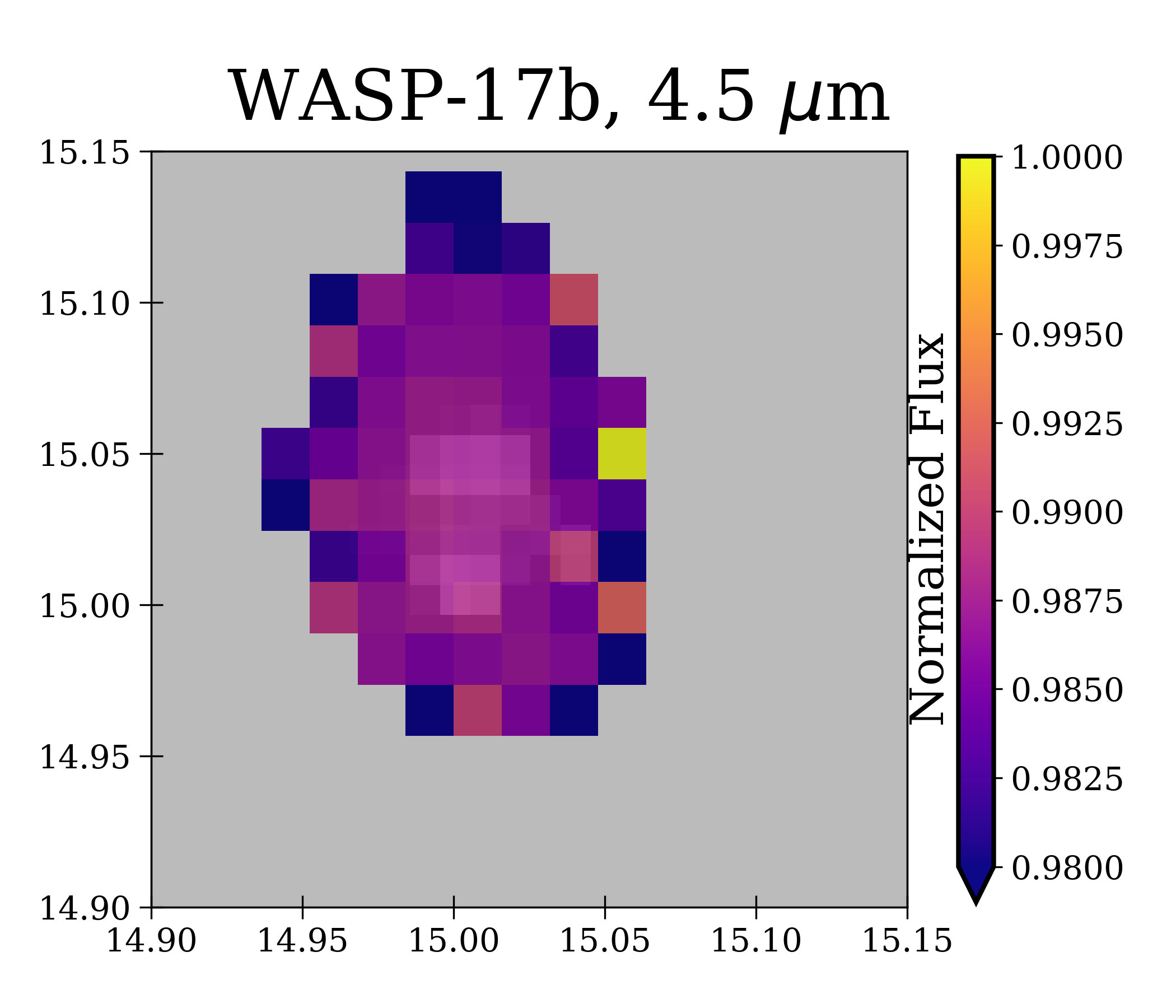}
    \caption{\textbf{Top:} Standard BLISS map best fit for the 3.6\,$\micron$ transit. \textbf{Bottom:} Fixed sensitivity map for the 4.5\,$\micron$ transit. The centroids for this transit overlap entirely with the fixed sensitivity map.}
    \label{fig:SPITZERbliss}
\end{figure}

\begin{figure*}
    \centering
    \includegraphics[width=\linewidth]{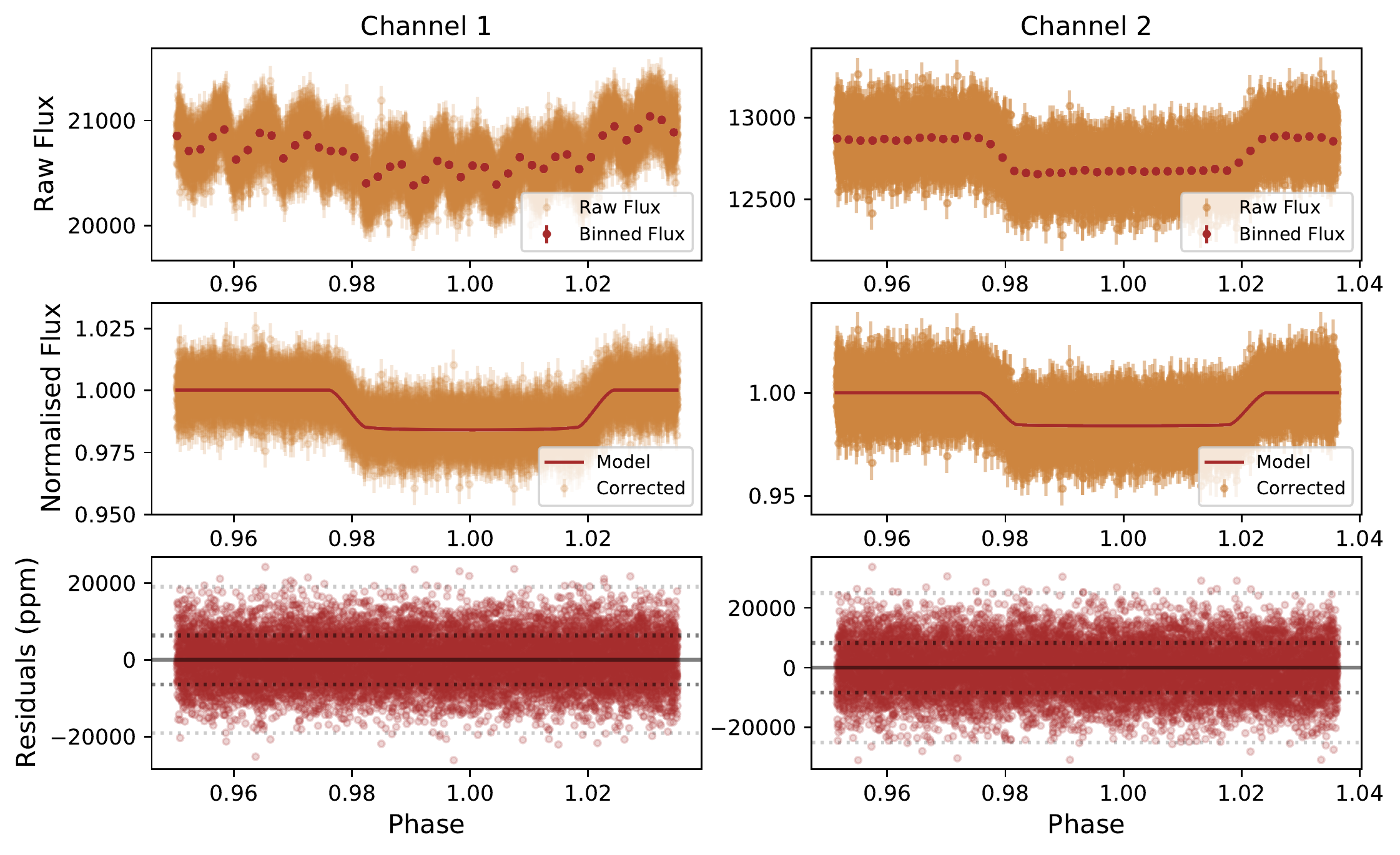}
    \caption{Photometric transit light curves of WASP-17b for two \textit{Spitzer} IRAC observations, Channel 1 (left) and Channel 2 (right). Top: The raw light curves along with the flux binned to phase increments of 0.002 (645\,$s$). Middle: Corrected, normalised light curves and best fit model. Bottom: Corrected light curve residuals. The dashed lines indicate one and three times the standard deviation.}
    \label{fig:spitzer_wlcs}
\end{figure*}

\subsubsection{The Intrapixel Sensitivity Effect}

Intrapixel sensitivity variations as the centroid drifts within a single pixel dominate the sources of \textit{Spitzer} IRAC systematics at 3.6 and 4.5\,$\micron$. To address this, we use the standard Bilinearly Interpolated Subpixel Sensitivity (BLISS) mapping technique introduced by \cite{Stevenson2012} at 3.6\,$\micron$, and the fixed sensitivity map introduced by \cite{May2020} at 4.5\,$\micron$. We use the standard BLISS mapping technique at 3.6\,$\micron$ instead of a fixed sensitivity map due to the time variability of the intrapixel effect in this channel, as discussed in \citet{May2020}.

The standard BLISS map is defined by (1) an intrapixel spatial binning size, optimised by comparing the Bayesian Information Criterion \citep[BIC,][]{Liddle2008} of map fits to those done with a nearest neighbour approach to ensure the data is not being overfit; and (2) the minimum number of exposure centroids required in a given spatial bin, optimised by comparing the standard deviation of the normalised residuals (SDNR) of fits varying this parameter. At 3.6\,$\micron$, which uses this standard BLISS technique, we used a spatial binning size of 0.006 pixels with a minimum number of 6 exposures in a given spatial bin. Spatial bins with less than 6 exposures are ignored in the fit and extrapolated over later.

The fixed sensitivity map needs only the spatial binning size optimised (See Figure \ref{fig:SPITZERbliss}). Here we selected the spatial binning size which results in the best BIC using the fixed map. At 4.5\,$\micron$, which uses the fixed sensitivity map, we used a bin size of 0.018 pixels. Because of the way this map is generated, there is no requirement on the number of exposures in a given spatial bin. The larger bin size at 4.5\,$\micron$ as compared to 3.6\,$\micron$ is reflective of the weaker intrapixel effect in this channel.

Figure \ref{fig:SPITZERbliss} shows the BLISS maps from each transit on the spatial bin size used.

\subsubsection{PRF Detrending}
At 3.6\,$\micron$ we also removed a functional dependence on the shape of the IRAC point-response function (PRF). The PRF stretches away from a circle towards an oval as the centroid approaches the edge of a pixel, resulting in lost flux when circular photometry is performed. There are demonstrated significant improvements in the quality of fits by removing this additional source of noise \citep{Lanotte2014,Demory2016a,Demory2016b,Gillon2017,Mendonca2018}. We find that a first order function of the PRFs performs best for this 3.6\,$\micron$ data. At 4.5\,$\micron$, this effect is encapsulated in the fixed sensitivity map.

\subsubsection{Astrophysical Source Models}
We used the python package \texttt{batman} \citep{Kreidberg2015} to model the Spitzer transit events. The orbital parameters are held fixed to those given in Table \ref{tab:lc_params}, fitting only for the centre of transit time and transit depth. Limb darkening is accounted for using a quadratic law based on the parameters given in Table \ref{tab:lc_params}, using \texttt{ExoCTK}'s limb darkening tool \citep{exoctk2021}. We adopted fixed values of \{0.069,0.127\} at 3.6\,$\micron$ and \{0.074, 0.091\} at 4.5\,$\micron$. We considered five possible temporal ramps: no ramp, an exponential ramp, a linear ramp, a combined exponential + linear ramp, and a quadratic ramp. The choice of ramp for each transit is identified by comparing the $\Delta$BIC value, with best fits calculated using a Levenberg-Marquardt minimiser. Parameter uncertainties are estimated using POET's custom Differential Evolution Markov Chain algorithm \citep{terBraak2006}. 

As discussed in \citet{Fu2021}, \textit{Spitzer} transit and eclipse depths can vary strongly with the amount of data trimmed from the start due to strong initial ramps, and due to temporal ramp modelling choices being degenerate with the standard BLISS map. To mitigate this effect, we progressively trim each dataset in 10 minute increments from 0 to 120 minutes, trying all 5 ramp options for each case. We find that the 3.6\,$\micron$ transit is not affected by this dependence, with the linear ramp always preferred, and consistent best fit transit depths across all trim levels for a linear ramp. We therefore select a 40 minute trim based on visually inspecting the data for our final fit. Due to separate degeneracies between BLISS mapping and PRF detrending that worsen if data are temporally binned \citep{May2020}, we do not perform temporal binning. 

Our best model combination at 3.6\,$\micron$ includes a linear temporal ramp, a first order PRF detrending function, a standard BLISS map, and a transit model. At 4.5\,$\micron$ we find that no ramp is needed, and our best model combination includes only the transit model and removing the fixed sensitivity map. The raw and resulting best fit \textit{Spitzer} photometric light curves are shown in Figure \ref{fig:spitzer_wlcs}, while the resulting transit depths are given in Table \ref{tab:wlc_depths}.

\section{The Panchromatic Transmission Spectrum of WASP-17b} \label{section:results}

\begin{figure*}
    \centering
    \includegraphics[width=\textwidth]{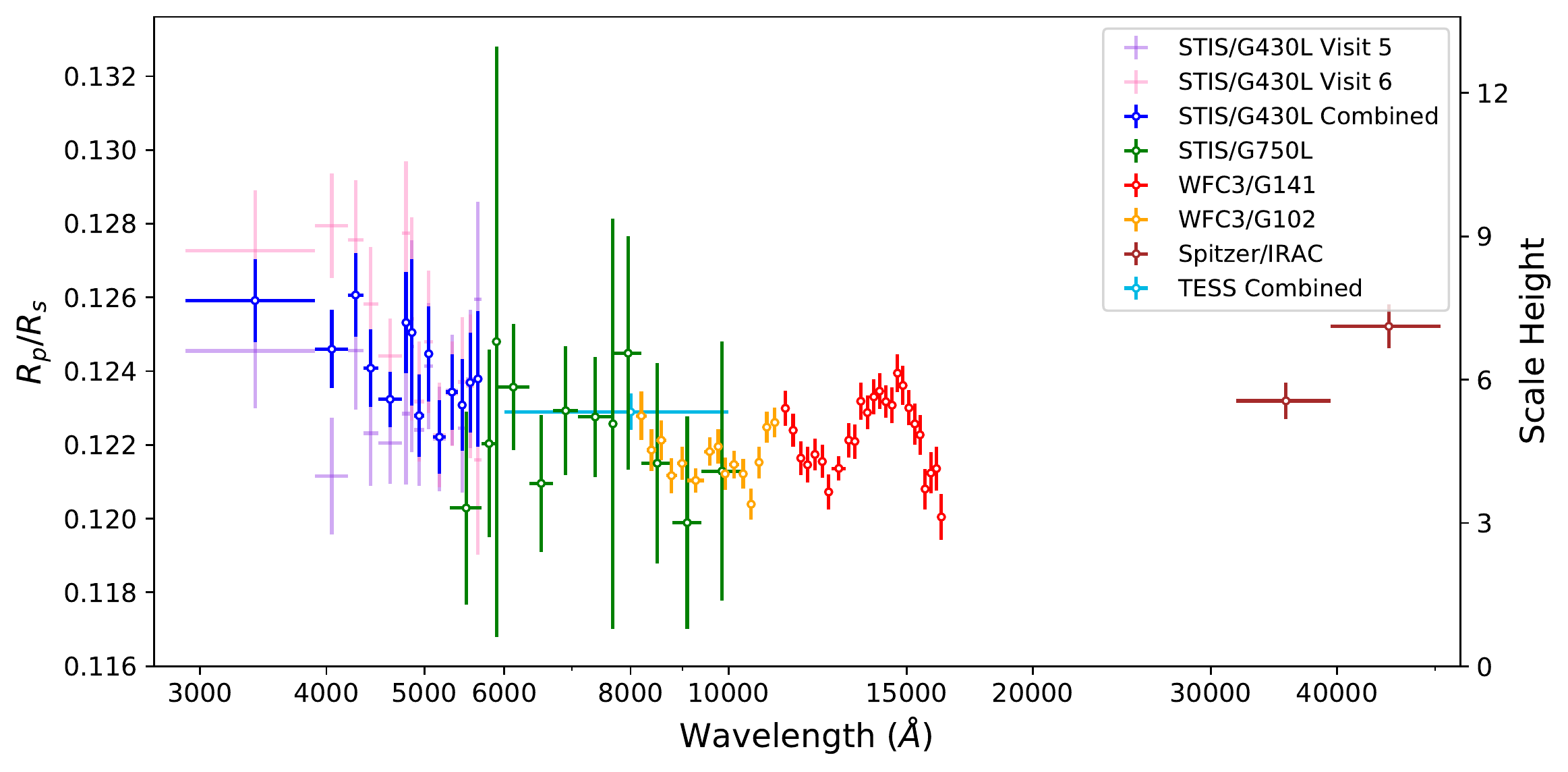}
    \caption{Transmission spectrum of WASP-17b measured over five visits with HST WFC3/G102 and G141, and STIS/G430L and G750L modes, and two visits with \textit{Spitzer} IRAC. The two individually analysed STIS/G430L visits are shown in pink and purple, and the transmission spectrum resulting from their weighted average is shown by the blue circles, spanning 3400--5650\,\AA. The transmission spectrum obtained by STIS/G750L from 5500--9850\,\AA\, is shown in green. The WFC3/G102 spectrum is shown in orange, and covers 8199--11110\,\AA, while the WFC3 G141 spectrum is shown in red, and covers 11380--16238\,\AA. The photometric observations by \textit{Spitzer} IRAC Channels 1 and 2 are shown in brown. Also plotted is the transit depth obtained from the joint fit of the transits observed by TESS, shown in turquoise. The scale height of WASP-17b was calculated using parameters obtained from \citet{Anderson2011}, assuming an H/He atmosphere with a mean molecular mass of $\mu=2.3$.}
    \label{fig:transmission}
\end{figure*}

Our 0.3--5.0\,\micron\, transmission spectrum of WASP-17b is shown in Figure \ref{fig:transmission}, wherein no offsets have been applied between the datasets. We take the weighted mean of the two G430L transmission spectra to produce the final quoted transmission spectrum between 0.340-0.565\,\micron. The spectrum is characterised by distinct H$_2$O absorption features at 1.15 and 1.4\,\micron, a weak slope in the blue-optical and a lack of evidence for absorption from \ion{Na}{i} and \ion{K}{i} (see Section \ref{section:iic}). In addition, our \textit{Spitzer} observations show WASP-17b to have a deeper transit at 4.5\,\micron\, than at 3.6\,\micron, indicative of absorption by carbon bearing species such as CO or CO$_2$ /citep{Fortney2010}. To ensure that our transmission spectrum is robust regardless of the choice of binning scheme, we also test a variety of bin sizes and positions across the STIS and WFC3 wavelength ranges, and found that in all cases, the resulting transmission spectra are consistent within 1 sigma, showing the same shape and structure. 

\subsection{Search for Sodium and Potassium} \label{section:iic}

\begin{figure}
    \centering
    \includegraphics[width=\linewidth]{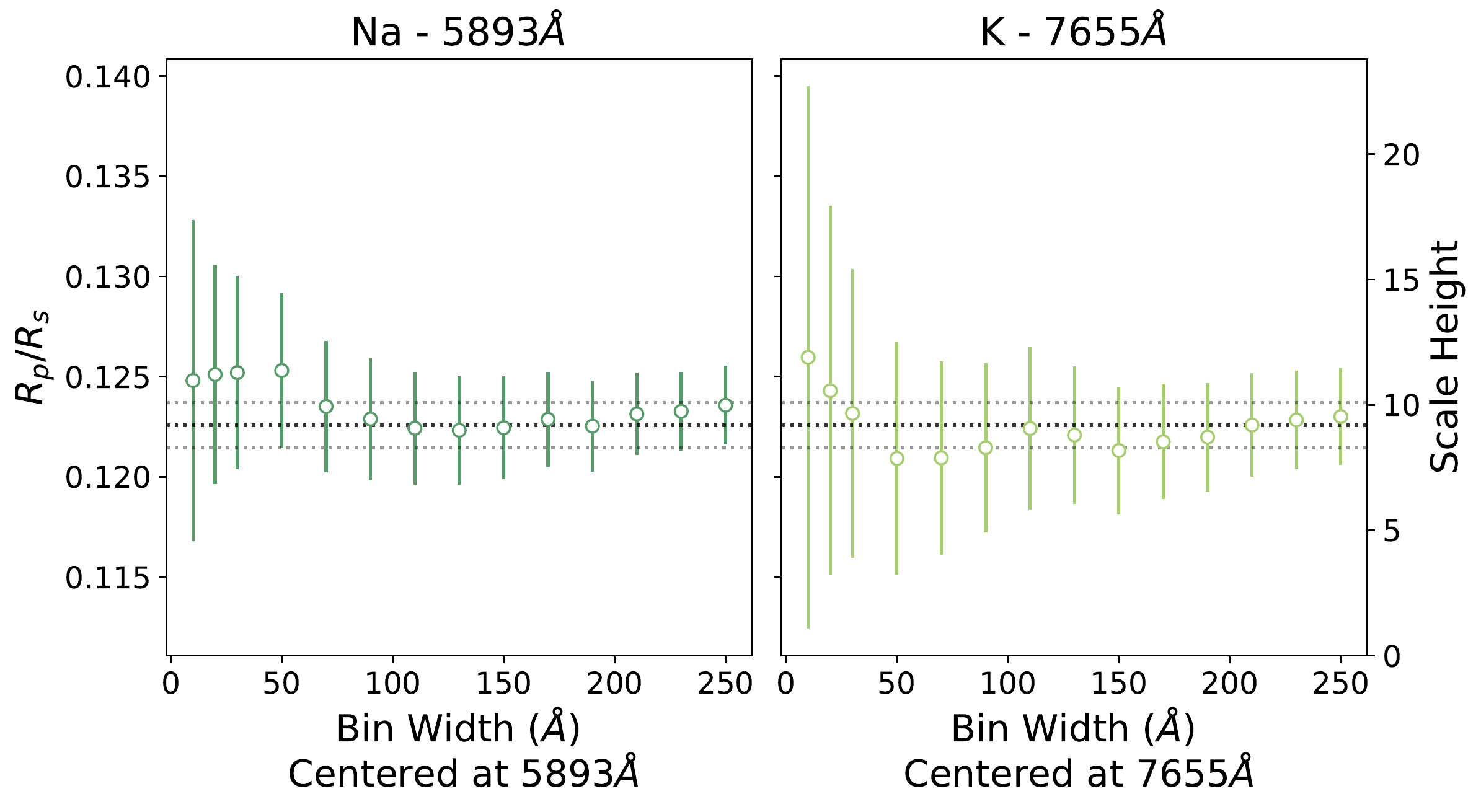}
    \caption{Left: \ion{Na}{i} absorption feature at 5893\,\AA\, shown in bins of incrementally increasing width from 10--250\,\AA. Right: \ion{K}{i} absorption feature at 7655\,\AA, shown in bins of incrementally increasing width from 10--250\,\AA. In both plots, dashed lines indicate the transit depth and one sigma uncertainties of the white light curve from the G750L observations (see Figure \ref{fig:stis_wlcs}). Both plots show a lack of increased absorption in narrower bins, indicating a non-detection of the respective features.}
    \label{fig:iic}
\end{figure}

We investigate the presence of \ion{Na}{i} and \ion{K}{i} absorption features in the G750L spectrum, as the original analysis of the G750L spectrum by \citet{Sing2016} showed evidence of \ion{Na}{i} absorption, but not of \ion{K}{i}. We make use of spectroscopic channels with a range of widths, increasing incrementally in steps of 10\,\AA\, from 10 to 30 \,\AA\,, and then in steps of 20\,\AA\, from 30 to 250\,\AA\, with each bin centred on 5893\,\AA\, for \ion{Na}{i} and 7665\,\AA\, for \ion{K}{i}. If an atmospheric signal is present in the planetary spectrum, this binning scheme should show a gradual decay in the transit depth with increasing bin width \citep[e.g.,][]{Alam2018, Alam2020, Alderson2020}. The results of this analysis are shown in Figure \ref{fig:iic}. Both the \ion{Na}{i} and \ion{K}{i} tests show a very small increased depth in the narrowest bin, but are well within the 1$\sigma$ uncertainty of the white light depth, indicating non-detections of the respective features. 

\subsection{Comparison to Previous Work} \label{section:oldstuff}

\begin{figure}
    \centering
    \includegraphics[width=\linewidth]{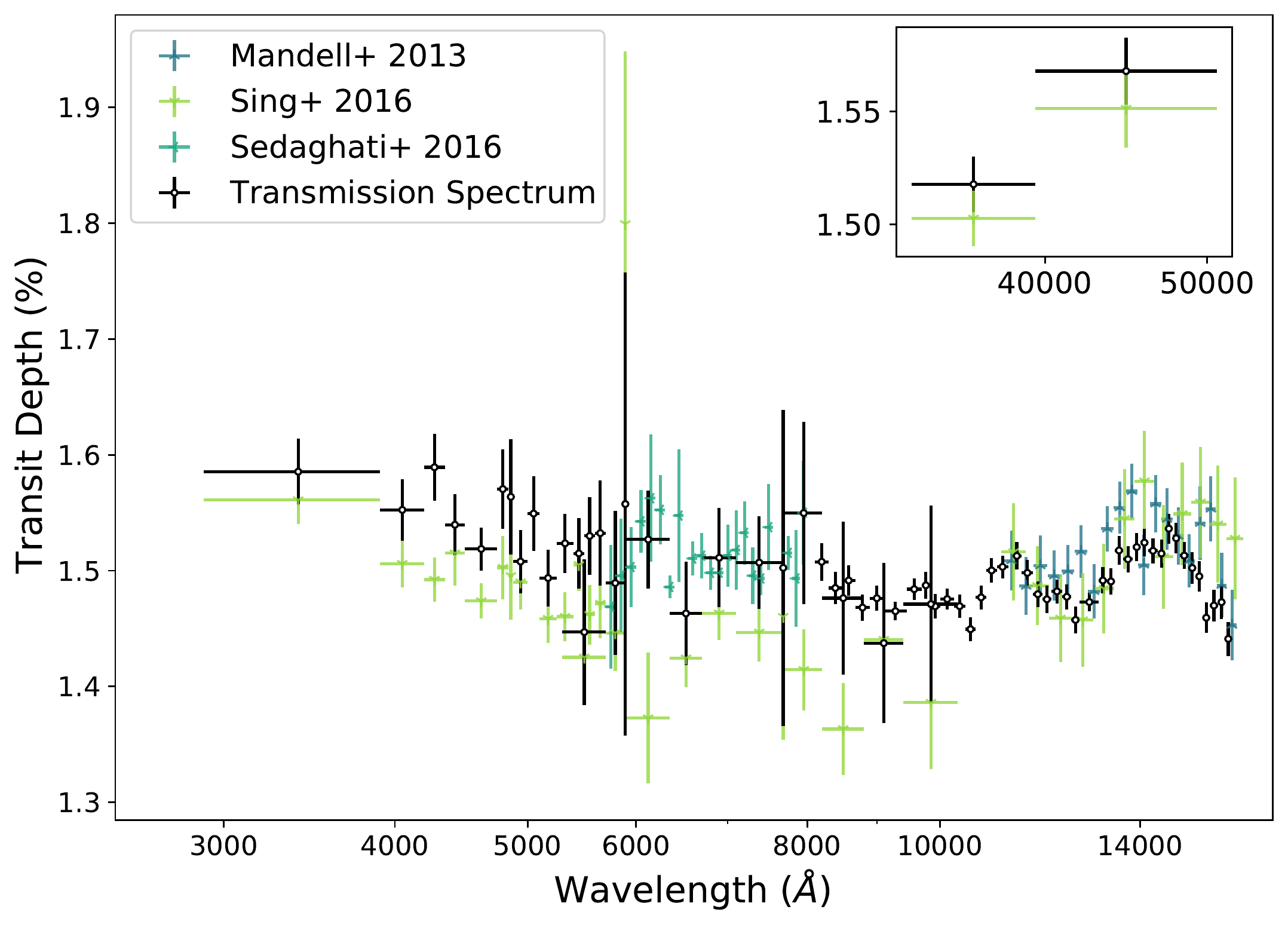}
    \caption{The transmission spectra of WASP-17b as presented by this work (black), along with that of \citet{Mandell2013} (blue), \citet{Sing2016} (lime green), and \citet{Sedaghati2016} (turquoise).}
    \label{fig:transmission_others}
\end{figure}

In addition to our HST+\textit{Spitzer} transmission spectrum, there also exists ground based transmission spectroscopy, presented by \citet{Sedaghati2016}, the original STIS and \textit{Spitzer} analysis by \citet{Sing2016}, and prior WFC3/G141 observations analysed by both \citet{Mandell2013} and \citet{Sing2016}. We show a comparison between these spectra and that of this work in Figure \ref{fig:transmission_others}. The shape of our transmission spectrum is broadly consistent with those of \citet{Sedaghati2016}, \citet{Sing2016} and \citet{Mandell2013}. Although there is an offset between the spectra presented in this work and by \citet{Sing2016}, this is due to the differences in the system parameters used in the analyses (see Table \ref{tab:lc_params} for updated and consistent system parameters). 

Notably, our new WFC3/G141 measurements have a greater precision compared to previous analyses, which have an average uncertainty of 250\,ppm over 19 bins in the case of \citet{Mandell2013} and 393\,ppm over 14 bins for \citet{Sing2016}, compared to the 120\,ppm average uncertainty over 25 bins achieved in this work. This marked improvement is driven by the implementation of WFC3's spatial scan mode and the more complete transit coverage. The addition of WFC3/G102 data is also beneficial, bridging the gap in wavelength coverage bluewards of WFC3/G141. This new overlap between WFC3/G102 and the red end of STIS/G750L aides in anchoring the STIS data in the wavelength region where fringing is known to have an impact, verifying the depth calibration of the transmission spectrum across instruments and allowing for the detection of multiple H$_2$O absorption peaks.

In the optical regime, we utilise the same binning scheme as that of \citet{Sing2016}, but have the added advantage of using consistent and more robust planetary parameters across all instruments, helping to avoid offsets between the discontinuous datasets from each visit. Our non-detection of \ion{Na}{i} absorption (see Section \ref{section:iic}) matches that of \citet{Sedaghati2016}. Additionally, while \citet{Wood2011} and \citet{Zhou2012} were able to detect an excess transit signal at the wavelength of \ion{Na}{i} in narrow 1.5\,\AA\, bins at $4\sigma$ and $4.5\sigma$ confidence respectively, \citet{Wood2011} found no significant detection in bins wider than 4\,\AA, consistent with our result. However, unlike \citet{Sedaghati2016}, we also find no evidence of \ion{K}{i} absorption. This can likely be explained by the fact that the ground-based spectroscopy of \citet{Sedaghati2016} was unable to probe the line core of \ion{K}{i}, so the quoted detection was driven only by potential evidence of the wings of the feature which would increase the continuum level of the spectrum. Figure \ref{fig:iic} shows that while the \ion{K}{i} absorption signal does appear to drop away in larger bin sizes, suggestive of wings around the line core, the change is fully encompassed by our observational uncertainties. Therefore, any detection of \ion{K}{i} remains inconclusive.


To contextualise the amplitude of the 1.4\,$\micron$\, H$_2$O absorption feature in the measured transmission spectrum of WASP-17b, we calculated the H$_2$O -- J index \citep{Stevenson2016} and the H$_2$O absorption amplitude \citep{Wakeford2019}. Using our transmission spectrum as presented in Figure \ref{fig:transmission}, we find a H$_2$O -- J index of $1.10 \pm 0.18$. This is a higher amplitude and more precise value than obtained in previous works (\citealt{Stevenson2016, Fu2017}, using the transmission spectrum of \citealt{Mandell2013}), and is thanks to the more precise G141 data provided by our use of spatial scan mode observations and improvements in transit light curve analysis techniques. The H$_2$O absorption amplitude is $35.8 \pm 3.4$\,\% muted compared to that of a clear solar metallicity atmosphere, which places WASP-17b firmly within the prediction that H$_2$O absorption features will typically be muted by aerosol opacities to $33\pm24$\,\% \citep{Wakeford2019}.

\section{Interpretation of the Atmosphere of WASP-17b} \label{section:atmosphere}

\begin{table}
\centering
\caption{Uniform prior bounds applied to the \texttt{POSEIDON} retrieval suite.} 
\label{tab:poseidon_posteriors}
\begin{tabular}{cc} 
\hline
                        & POSEIDON Priors \\ 
\hline
$R_\mathrm{p, ref}$ ($R_J$)           & 1.5895, 2.1505 \\
$T$ (K)                     & 400, 2300 \\
log($X_\mathrm{i}$)    & -12, -1 \\
log($a$)                & -4, 8   \\
$\gamma$                & -20, 2  \\
log($P_\mathrm{cloud}$) & -6, 2   \\
\hline
\end{tabular}
\centering
\end{table}

\begin{table*}
    \centering
    \caption{Goodness of fit values and major species abundances from the suite of free chemistry retrievals run on the WASP-17b transmission spectrum (see Figure \ref{fig:retrieval_stats}). The full transmission spectrum consists of 68 data points such that the degrees of freedom represents 68 - the number of free parameters. Each retrieval listed includes: three aerosol parameterisations ($\alpha$, $\gamma$, log($P_\mathrm{cloud}$)) and an isothermal temperature profile. Model 19 represents the most basic model. Bolded models are those presented in Figure \ref{fig:retrieval}.}
    \begin{tabular}{l|l|cccccc}
         Model &  Fit chemical species & DOF &  $\chi^2$ CDF (\%) & BIC & ln($Z$) & log($X_\mathrm{H_2O}$) & log($X_\mathrm{CO_2}$) \\
         \hline
        19 & H$_2$O, CH$_4$, CO$_2$ & 60 & 73.96 & 100.36 & 469.17 & -4.90$^{+0.43}_{-0.27}$ & -5.80$^{+0.72}_{-0.74}$ \\
        18 & H$_2$O, CH$_4$, CO$_2$, CO & 59 & 78.47 & 105.24 &	469.02 & -4.86$^{+0.45}_{-0.29}$  & -5.83$^{+0.74}_{-0.80}$ 	\\
      \textbf{17 (B)}  & \textbf{H$_2$O, CH$_4$, CO$_2$, Na, K, CO} & \textbf{57} & \textbf{70.15} & \textbf{108.54} & \textbf{468.94} & \textbf{-4.73$^{+0.65}_{-0.32}$} & \textbf{-5.64$^{+0.89}_{-0.81}$} \\
        16 & H$_2$O, CH$_4$, CO$_2$, TiO & 59 & 73.82 & 103.47 & 468.56 & -4.69$^{+0.68}_{-0.35}$ & -5.55$^{+0.90}_{-0.78}$ \\
        15 & H$_2$O, CH$_4$, CO$_2$, Na, K, CO, TiO & 56 & 73.38 & 112.79 & 468.22 & -4.44$^{+1.60}_{-0.48}$ & -5.25$^{+1.52}_{-0.94}$ \\
        14 & H$_2$O, CH$_4$, CO$_2$, Na, K, CO, VO & 56 & 73.38 & 112.79 & 467.41 & -4.42$^{+1.97}_{-0.55}$ & -5.20$^{+1.88}_{-1.06}$ \\
        13 & H$_2$O, CH$_4$, CO$_2$, H$^-$ & 59 & 60.86 & 99.34 & 468.63 & -4.22$^{+1.92}_{-0.91}$  & -5.22$^{+2.00}_{-1.41}$ \\
      \textbf{12 (A)} & \textbf{H$_2$O, CH$_4$, CO$_2$, H$^-$, TiO} & \textbf{58} & \textbf{58.78} & \textbf{101.94} & \textbf{468.35} & \textbf{-3.21$^{+1.29}_{-1.62}$} & \textbf{-4.19$^{+1.43}_{-1.88}$} \\
        11 & H$_2$O, CH$_4$, CO$_2$, H$^-$, VO & 58 & 62.80 & 103.10 & 467.47 & -2.98$^{+1.09}_{-1.85}$ & -4.06$^{+1.30}_{-2.06}$ \\
        10 & H$_2$O, CH$_4$, CO$_2$, H$^-$, Na, K, CO & 56 & 60.71 & 108.87 & 468.80 & -2.91$^{+1.05}_{-1.83}$ & -3.85$^{+1.19}_{-2.13}$ \\
        9  & H$_2$O, CH$_4$, CO$_2$, H$^-$, Na, K, TiO & 56 & 62.67 & 109.43 & 468.66 & -2.32$^{+0.59}_{-1.77}$ & -3.29$^{+0.79}_{-1.80}$ \\
        8  & H$_2$O, CH$_4$, CO$_2$, H$^-$, Na, K, VO & 56 & 62.67 & 109.43 & 468.01 & -2.30$^{+0.89}_{-2.82}$ & -3.25$^{+1.63}_{-5.00}$ \\
        7  & H$_2$O, CH$_4$, CO$_2$, H$^-$, Na, K, CO, TiO & 55 & 66.35 & 113.70 & 468.60 & -2.28$^{+0.58}_{-1.69}$ & -3.25$^{+0.76}_{-1.79}$ \\     
        6  & H$_2$O, CH$_4$, CO$_2$, H$^-$, Na, K, CO, VO & 55 & 64.50 & 113.15 & 468.28 & -2.16$^{+0.49}_{-1.49}$ & -3.13$^{+0.69}_{-1.58}$ \\
         \textbf{5 (C)}    & \textbf{H$_2$O, CH$_4$, CO$_2$, H$^-$, Na, K, CO, VO, TiO} & \textbf{54} & \textbf{68.04} & \textbf{117.39} & \textbf{467.97} & \textbf{-1.99$^{+0.41}_{-0.98}$} & \textbf{-2.94$^{+0.58}_{-1.06}$}\\
        4  & H$_2$O, CH$_4$, CO$_2$, H$^-$, Na, K, CO, VO, CrH, FeH & 53 & 72.94 & 122.12 & 466.55 & -1.87$^{+0.34}_{-0.59}$ & -2.80$^{+0.51}_{-0.74}$ \\
        3  & H$_2$O, CH$_4$, CO$_2$, H$^-$, Na, K, CO, TiO, VO, CrH, FeH & 52 & 75.83 & 126.27 & 466.32 & -1.80$^{+0.31}_{-0.49}$ & -2.72$^{+0.46}_{-0.64}$\\
        2  & H$_2$O, CH$_4$, CO$_2$, H$^-$, Na, K, CO, TiO, VO, CrH, FeH, & 50 & 82.04 & 134.95 & 465.31 & -1.69$^{+0.27}_{-0.36}$ & -2.60$^{+0.42}_{-0.53}$ \\
         & AlO, TiH \\
        1  & H$_2$O, CH$_4$, CO$_2$, H$^-$, Na, K, CO, TiO, VO, CrH, FeH, & 43 & 93.76 & 163.54 & 465.98 & -1.68$^{+0.26}_{-0.31}$ & -2.59$^{+0.43}_{-0.49}$\\
              &  \ \  AlO, TiH, Li, SiO, AlH, CaH, SiH, HCN, SH (No CH$_4$) & \\
        \hline
        A IR-Only & H$_2$O, CH$_4$, CO$_2$, H$^-$, TiO & 32 & 96.32 & 89.88 & 300.89 & -2.75$^{+0.80}_{-1.24}$ & -3.81$^{+0.98}_{-1.43}$ \\
        B IR-Only & H$_2$O, CH$_4$, CO$_2$, Na, K, CO & 31 & 98.29 & 96.32 & 300.82 & -2.23$^{+0.47}_{-0.87}$ & -3.22$^{+0.65}_{-1.00}$ \\
        C IR-Only & H$_2$O, CH$_4$, CO$_2$, H$^-$, Na, K, CO, VO, TiO & 28 & 98.22 & 104.99 & 300.48 & -3.93$^{+1.20}_{-0.68}$ & -4.81$^{+1.21}_{-0.96}$ \\
        \hline
        \hline
        ATMO &  Fixed [M/H], Includes opacities from: \\ 
        & (H$_2$O, CH$_4$, CO$_2$, H$^-$, Na, K, CO, TiO, VO, & 62 & 77.19 & 95.27 & - & -2.10$^{+0.22}_{-0.78}$ & -3.93$^{+0.41}_{-0.85}$ \\
                & \ \ FeH, Li,Rb, Cs, PH$_3$, H$_2$S, HCN, C$_2$H$_2$, SO$_2$, Fe, NH$_3$) & \\
        \hline
    \end{tabular}
    \label{tab:retrieval_stats}
\end{table*}

\begin{figure*}
    \centering
    \includegraphics[width=\linewidth]{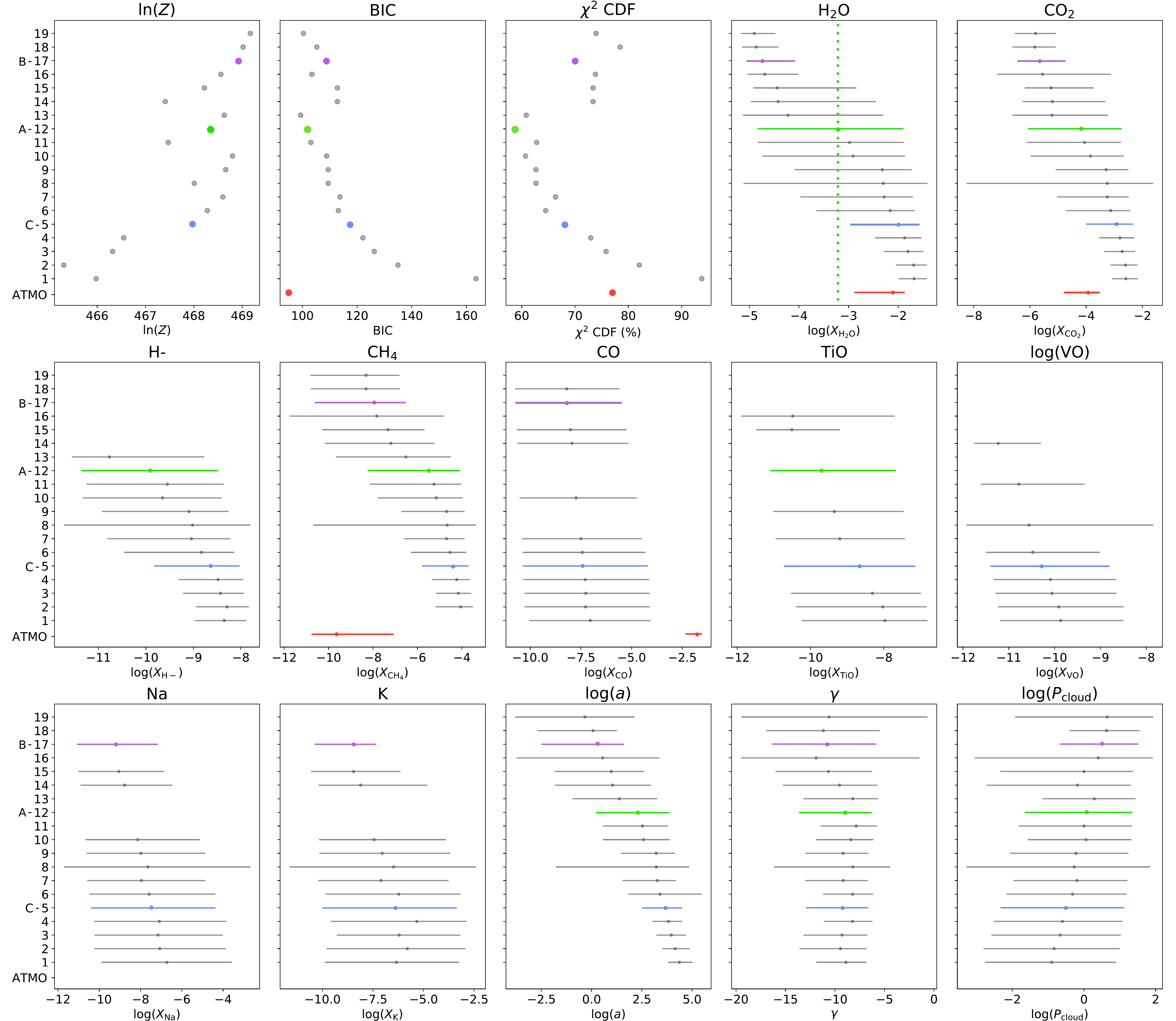}
    \caption{Retrieval statistics, retrieved parameter values and 1 $\sigma$ uncertainties for the suite of \texttt{POSEIDON} free chemistry retrievals (grey points). Each retrieval is labelled 1 - 19 in order of H$_2$O abundance from high to low values. The retrieval preferred by reducing the $\chi^2$ CDF, 12, is shown in green (referred to as Model A in the text and Figure \ref{fig:retrieval}). The purple (17, B) and blue (5, C) points show retrievals from the low and high abundance modes of H$_2$O which have the best $\chi^2$ CDF. Retrievals B and C are used throughout the text and in Figure \ref{fig:retrieval} to highlight the differences between the low and high abundance modes of H$_2$O, the separation between which is shown by the vertical green line in the H$_2$O panel. Results from the \texttt{ATMO} equilibrium chemistry retrieval are shown by the red points. The ln($Z$) for \texttt{ATMO} is not shown as due to the differences in the retrieval two codes, the resulting values are not comparable.}
    \label{fig:retrieval_stats}
\end{figure*}

\begin{figure*}
    \centering
    \includegraphics[width=\linewidth]{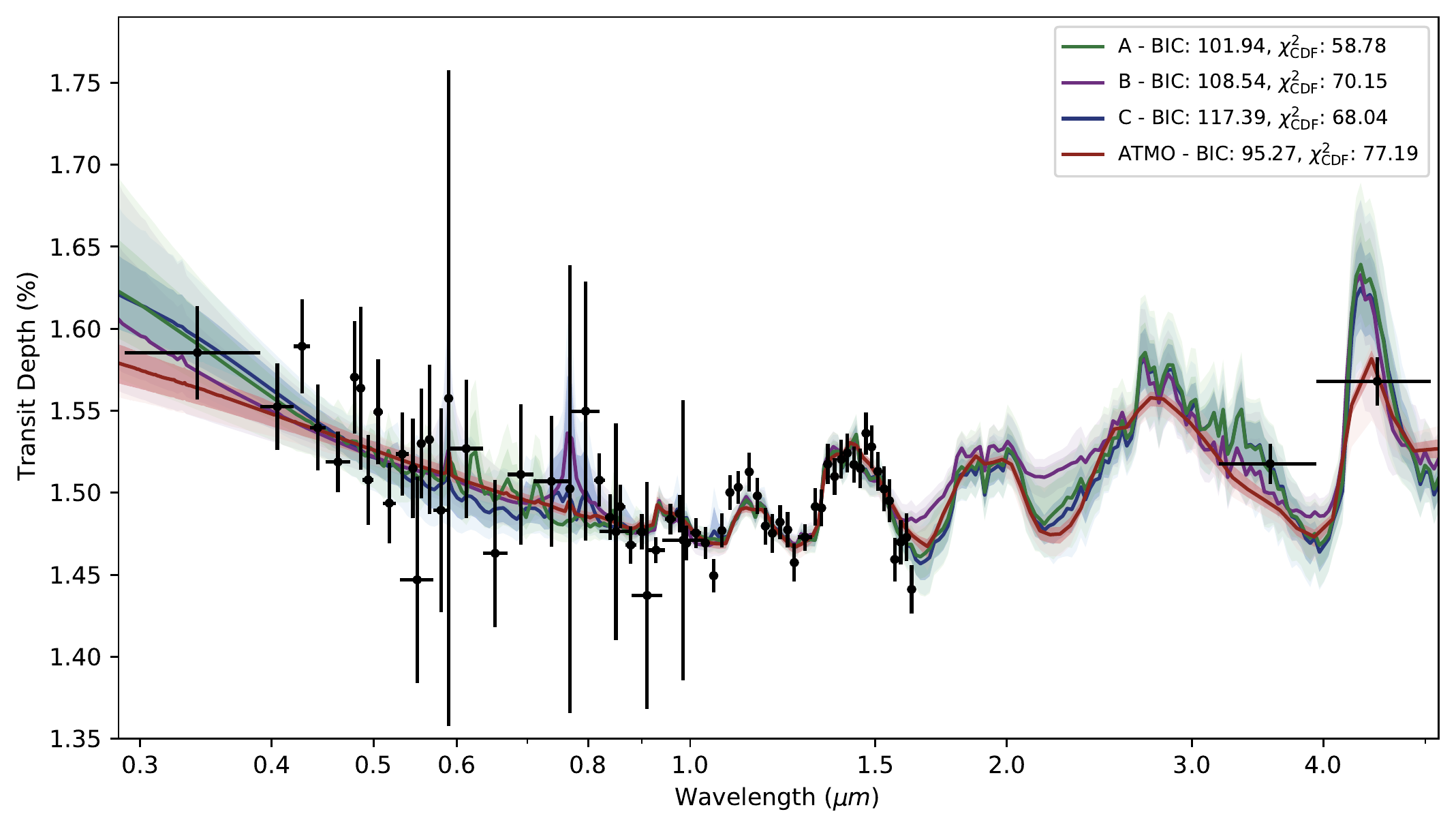}
    \includegraphics[width=\linewidth]{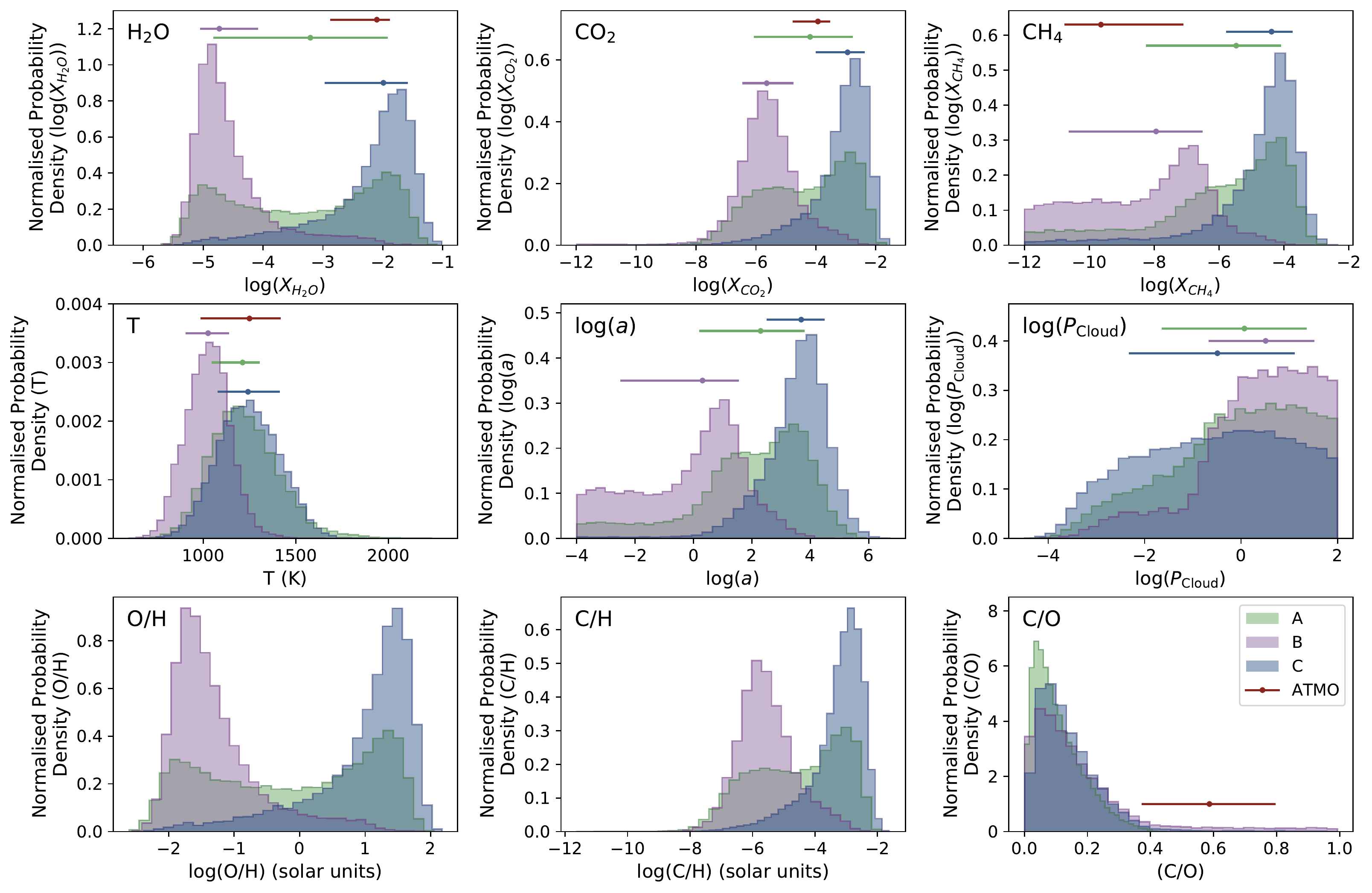}
    \caption{Top: Retrieved transmission spectra for the three highlighted  \texttt{POSEIDON} free chemistry retrievals (Model A in green, Model B in purple and Model C in blue), and for the \texttt{ATMO} equilibrium chemistry retrieval (in red). For each model, the 1 and $2\sigma$ uncertainties are given by the shaded regions. The transmission spectrum of WASP-17b as obtained by HST and \textit{Spitzer} is shown by the black points. Bottom: Marginalised posterior probability distribution histograms of the three highlighted \texttt{POSEIDON} free chemistry retrievals (Model A in green, Model B in purple and Model C in blue) for a selection of retrieved parameters, along with the inferred O/H, C/H and C/O ratios (derived from the mixing ratio samples). In the case of retrieved parameters, errorbars show the median retrieved values and the $1\sigma$ confidence level. Retrieved parameters from \texttt{ATMO} are shown as red errorbars at the top of the relevant histogram plots.}
    \label{fig:retrieval}
\end{figure*}

To interpret the measured transmission spectrum of WASP-17b, we ran a suite of atmospheric retrieval exercises to better understand the thermochemical properties and compositions that contribute to the atmosphere. Atmospheric retrieval analysis involves the fitting of atmospheric models to an observed spectrum via a parameter estimation technique in order to obtain robust inferences of the uncertainties and trends in the parameters \citep{Madhu2009}. Retrieval frameworks can operate using a free chemistry model (e.g., POSEIDON \citet{MacDonald2017}), in which the volume mixing ratios of each chemical species included within a model atmosphere are treated as separate free parameters, or using a chemical equilibrium model (e.g., ATMO \citet{Amundsen2014, Tremblin2015, Wakeford2017Sci}), in which the abundances of the complete chemical inventory are solved for assuming that the chemical processes governing the atmosphere are in equilibrium. By utilising multiple retrieval methods, the impacts of different modelling choices can be quantified, while comparison between them can hint at processes that would be difficult to detect with a single method \citep[e.g.,][]{Baudino2017,Barstow2018,Lewis2020, Rathcke2021,Barstow2022}. For example, comparing the results of free and equilibrium chemistry retrievals can help to ascertain whether disequilibrium processes are at play \citep[e.g.,][]{Baudino2017,Lewis2020}. We therefore employ a suite of free chemistry retrievals which we compare to the results of an equilibrium chemistry retrieval (see Section \ref{section:atmo}).

We conducted a comprehensive suite of free chemistry retrievals using the POSEIDON retrieval framework \citep{MacDonald2017}. POSEIDON couples an exoplanet radiative transfer module with a Bayesian sampling algorithm to explore the range of atmospheric properties consistent with a given transmission spectrum. As POSEIDON operates under free chemistry, we use this framework to explore the combination of different chemical species and retrieval set-ups, and later compare it to equilibrium chemistry, in which the network of chemical species is fixed (see Section \ref{section:atmo}). We explore the parameter space using \verb|PyMultiNest| \citep{Feroz2008, Feroz2009, Buchner2014, Feroz2019}, a Bayesian nested sampling algorithm implemented in python, using 4,000 live points.

We investigated a broad range of scenarios to explain our WASP-17b transmission spectrum. Across our suite of free retrievals we consider an extensive list of optical and IR opacity sources, including: Na, K, Li, H$_2$O, CH$_4$, CO, CO$_2$, TiO, VO, AlO, SiO, TiH, CrH, FeH, AlH, CaH, SiH, HCN, and NH$_3$. Our cross sections are derived from ExoMol \citep{Tennyson2016} and VALD3 \citep{Pakhomov2017} line lists, as detailed in \citet{MacDonald_Thesis_2019}. We also include continuum opacity from H$^{-}$ \citep{John1988}, H$_2$-H$_2$ and H$_2$-He collision-induced absorption from HITRAN \citep{Karman2019}, and Rayleigh scattering. We describe aerosols via a cloud deck and a wavelength-dependent scattering haze, both of which may be inhomogenous around the terminator \citep{MacDonald2017}. The scattering is parameterised by log($a$), the Rayleigh enhancement factor, and $\gamma$, the scattering slope. The cloud deck is parameterised by log($P_\mathrm{cloud}$), representing the top pressure of a cloud within infinite opacity at all wavelengths \citep[see][for further details]{MacDonald2017}. We consider both an isothermal, gradient and a parametric pressure-temperature (P-T) profile \citep{Madhu2009}. The impacts of stellar contamination \citep{Rackham2017} on WASP-17b's transmission spectrum are parameterised by the temperature and covering fraction of stellar heterogeneities and the photosphere temperature \citep[see][]{Rathcke2021}. All retrievals also fit for the planetary radius at the 10 bar reference pressure, $R_{p, \, \rm{ref}}$. We compute our model spectra from 0.28 to 5.2\,$\micron$ at $R =$ 4,000 before binning down to the resolution of the observations.

Throughout our retrieval analysis, we employ three statistical metrics to determine the optimal retrieval, defined by a set of free parameters, from which we draw its corresponding best fitting parameters to define the transmission spectrum model. We make use of the maximisation of the evidence, ln($Z$), along with the minimisation of the BIC to define the optimised retrieval parameterisation. We also consider a frequentist metric given by the cumulative distribution function (CDF) of the $\chi^2$ distribution \citep[see][for a discussion of the $\chi^2$ CDF in a model selection context]{Wilson2021}, which defines the probability that a better fit to the most likely model in that retrieval can be obtained given a random draw of residuals with the same uncertainties.

\subsection{Determining Retrieval Model Complexity}\label{section:bulk}

We first conduct several exploratory retrievals to assess the necessary model complexity to fit WASP-17b's transmission spectrum. Our initial explorations of the transmission spectrum test three P-T profiles, an isotherm, a temperature gradient linear in log-pressure space, and the 6-parameter profile from \citet{Madhu2009}. The retrievals indicated that the transmitting atmosphere of WASP-17b is best modelled by an isothermal P-T profile, with a $\Delta$BIC of 16 and 26 compared to the gradient and \citet{Madhu2009} profiles respectively. These $\Delta$BIC values represent strong evidence against the need for the more complex P-T profiles.  

We next determined the minimum complexity cloud parameterisation required. We ran a series of retrievals with the same chemical species but varying the inclusion of one or both of uniform and wavelength dependent scattering cloud prescriptions. Our tests indicate the need to include both a cloud deck and a scattering parmeterisation, with a minimum 15\% improvement in the $\chi^2$ CDF compared to those with just one or without either prescription considered. We found that inhomogenous `patchy' clouds were disfavoured by the $\chi^2$ CDF by 10\% and $\Delta$BIC of 11, but did show a minimal improvement in the lnZ of $<$ 1, given these statistics and the current data quality we do not include these parameterisations in further retrieval fits shown in this paper. 

Finally, as stellar activity can mimic the presence of scattering slopes \citep{McCullough2014, Oshagh2014}, we additionally test the impacts of stellar contamination on the transmission spectrum from stellar spots and plages. The retrieved heterogeneity temperatures were fully consistent with the stellar photospheric temperature of WASP-17, with $<$10\% retrieved covering fractions, consistent with 0\% within the 2\,$\sigma$ uncertainties. This, combined with a log$R' _{\mathrm{HK}}$ value of -5.531 \citep{Sing2016}, and no prior evidence of significant stellar variability, suggests that stellar activity does not play a role in the observed transmission spectrum, and we therefore do not consider its inclusion further.

\subsection{Exploring the Chemical Inventory} \label{section:chemical_results}

Due to its 1770\,$K$\, equilibrium temperature, WASP-17b lies at a critical juncture within the close-in giant planet population, sitting between the ultra-hot and hot Jupiters at the traditional pM/pL boundary \citep{Fortney2008}, where atmospheres become warm enough for TiO and VO to have a significant opacity. Furthermore, P-T profiles by \citet{Kataria2016} show that the range of temperatures from the day- to the night-side of WASP-17b spans 1000-2500\,$K$, resulting in a wide variety of plausible spectroscopically active species. As a result of the potential chemical diversity of the atmosphere of WASP-17b, we explored a wide range of species within our \verb|POSEIDON| free-retrievals to evaluate the dominant absorbing species influencing WASP-17b's measured atmosphere. 

Across the suite of retrievals used to analyse the chemical composition we include scattering and uniform cloud parameterisations (three additional free parameters, log($a$), $\gamma$ and log($P_\mathrm{cloud}$)), an isothermal P-T profile (one free parameter, $T$), and opacities from a combination of chemical species, including H$^-$, Na, K, Li, H$_2$O, CH$_4$, CO, CO$_2$, TiO, VO, AlO, SiO, TiH, CrH, FeH, AlH, CaH, SiH, HCN, and NH$_3$. We summarise the priors for our free parameters in Table \ref{tab:poseidon_posteriors}. Our full exploration resulted in a suite of 19 retrievals that include absorption from a minimum of three spectroscopically active species (H$_2$O, CH$_4$ and CO$_2$) to a maximum of 20 (see Table \ref{tab:retrieval_stats}). The retrievals are labelled from 1 - 19, with retrieval 19 representing the most basic configuration and 1 the most complex, as shown in Figure \ref{fig:retrieval_stats}. 

We find that, based on minimising the $\chi^2$ CDF, the most statistically favoured model includes opacity sources from H$_2$O, CO$_2$, CH$_4$, H$^-$ and TiO, and is shown in Figure \ref{fig:retrieval}, labelled as A and plotted in green. Despite this retrieval seemingly being preferred, the posteriors for many parameters in this retrieval were bimodal or spanned a large parameter space, resulting in both high and low chemical abundances being possible, as seen in Figures \ref{fig:retrieval_stats} and \ref{fig:retrieval}. To explore the apparent degeneracies in the retrieval, resulting in a bimodal distribution in a number of key factors, we test a large number of other species combinations, to determine if a more physically viable solution could be retrieved (see Figure \ref{fig:retrieval}).

Within the suite, the best-fit models from each retrieval are ordered from the highest (Model 1) to the lowest (Model 19) retrieved H$_2$O abundance (see Figure \ref{fig:retrieval_stats}). We find a continuum of median H$_2$O abundances from -1.68 to -4.9, with multiple retrievals constraining abundances at either end of this range, and a selection of retrievals with bimodal distributions in between. As ordered by H$_2$O abundance in Figure \ref{fig:retrieval_stats}, our $\chi^2$ CDF preferred retrieval A is the 12th configuration. To compare the best fitting spectra obtained by the high and low abundance modes, we also highlight the retrieval with the best $\chi^2$ CDF which obtains a well constrained H$_2$O abundance for each mode. As shown in Figures \ref{fig:retrieval_stats} and \ref{fig:retrieval}, these are retrieval 17 for the low abundance mode and the retrieval 5 for the high abundance mode (henceforth referred to as B and C). The best fit models of A, B, and C show broad agreement at optical wavelengths, with small differences due to the range of opacity sources included in each retrieval. The three highlighted models diverge the most into the IR beyond 1.6\,$\micron$, with A and C fitting the downturn in the final four WFC3/G141 data points, and B producing a flatter spectrum after 1.6\,$\micron$ due to the difference in the more dominant species in each retrieval.

We see a broad range of trends in the retrieved parameters within our suite, with the abundances of CO$_2$ and CH$_4$ following the same pattern as that of H$_2$O. There is also a degeneracy between the chemical abundances and aerosol properties of the atmosphere, and trends in the inclusion of H$^-$. All retrievals find no evidence of Na or K where included. All three of the highlighted retrievals obtain consistent best fit values of $T$, and posteriors of the C/O ratios consistent with sub-solar values (see Figure \ref{fig:retrieval}). However, given the extent of bimodality within our retrieval suite, there is a lack of clarity as to the exact chemical composition of the atmosphere. Ordinarily, adding redundant parameters to the retrieval with the highest Bayesian evidence should not result in any significant changes in the conclusions of that retrieval. Such model instability likely speaks to poorly constraining data at key wavelengths, and we therefore must investigate the role of the data in these wavelength ranges.

\subsubsection{Inclusion of Optical Data} \label{section:optical}

To better explore the role of the optical data in the retrieved atmospheric abundances, we ran each of our three highlighted retrieval set-ups without the STIS/G430L and G750L optical data. In doing this, we are able to assess to what extent our chemical compositions and abundances are driven by the higher precision IR data alone, and are better able to understand the importance of the inclusion of optical data when constraining the impacts of uniform and wavelength-dependant aerosol scattering.

In all cases, we find that removing the STIS/G430L and G750L optical data from the retrievals results in worse $\chi^2$ CDF and leads to significantly poorer constraints on all three aerosol scattering parameters. Both the full transmission spectrum and IR-only retrievals achieve chemical abundances within $\sim1\sigma$ for the key species included in all retrievals (H$_2$O, CH$_4$ and CO$_2$). We also obtain consistent abundances and uncertainties for CO and the optical species considered in each retrieval, implying that the poorer precision achieved by STIS due to the partial transits observed (see Section \ref{section:reduction}) is a limiting factor in our ability to retrieve a better constrained inventory of the optical species present within the atmosphere. 

\begin{figure}
    \centering
    \includegraphics[width=\linewidth]{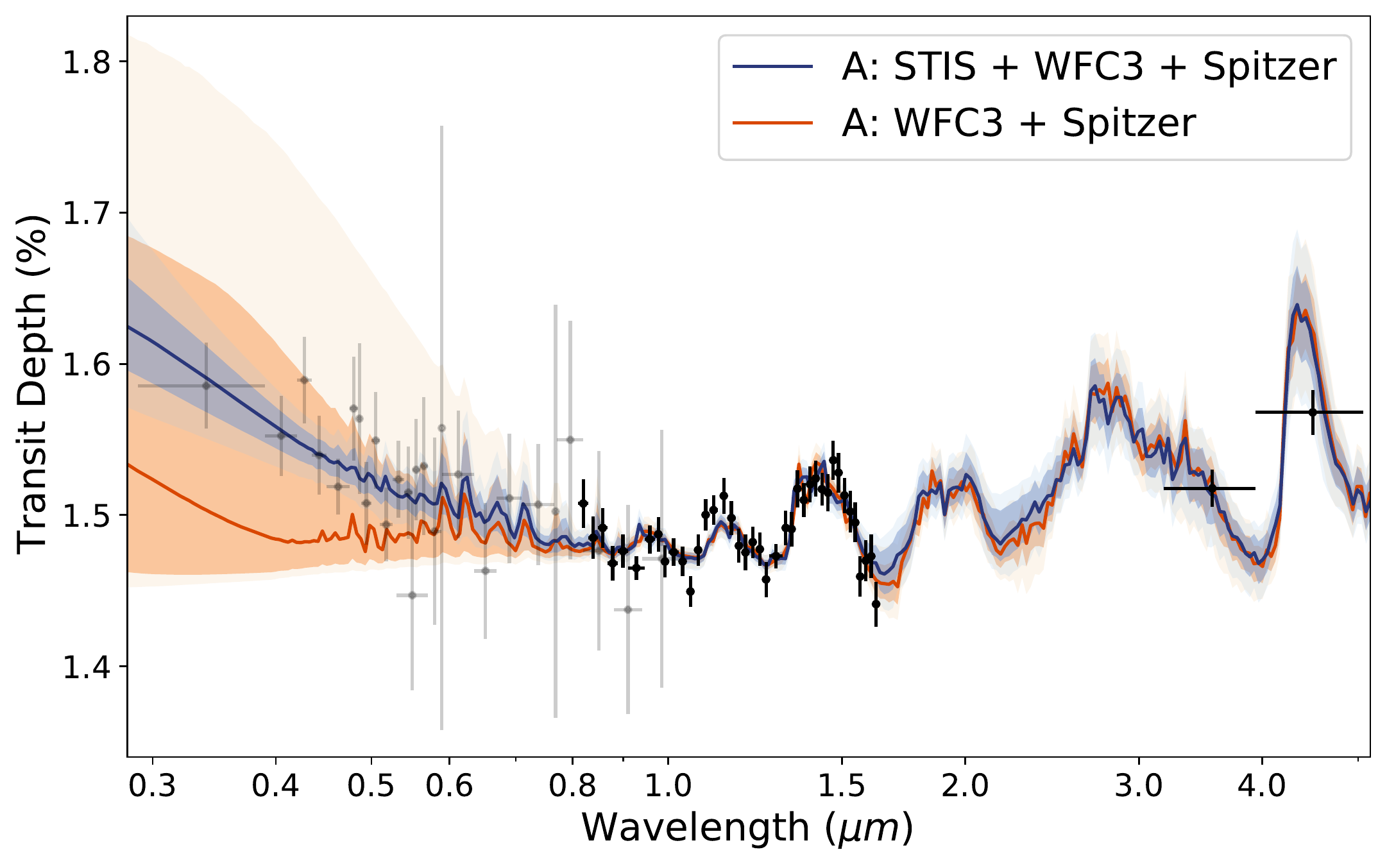}
    \caption{Comparison between the retrieved transmission spectra of a \texttt{POSEIDON} free chemistry retrieval of the full 0.3-5\,\micron\, transmission spectrum using STIS, WFC3/IR and \textit{Spitzer} (blue), and the of same retrieval without the inclusion of STIS data (orange). Both retrievals use the retrieval A configuration as outlined in Table \ref{tab:retrieval_stats}. For each model, the 1 and $2\sigma$ uncertainties are given by the shaded regions. The transmission spectrum of WASP-17b as obtained by WFC3/IR and \textit{Spitzer} is shown by the black points, while the STIS data is shown in grey. The difference in the $\sigma$ uncertainties between 0.3-0.8\,\micron\, highlights the importance of optical data in constraining the scattering slope properties of the atmosphere.}
    \label{fig:retrievalIR}
\end{figure}

We show a comparison between the full transmission spectrum and the IR-only retrieval for our statistically preferred configuration, retrieval A, in Figure \ref{fig:retrievalIR}. This highlights the similarity between the best fitting models at IR wavelengths, and also demonstrates the disparity at the optical wavelengths which drive constraints on aerosol scattering. Additionally, the removal of the optical data enables the retrieval to better converge on regions of the phase space as opposed to producing bimodal posterior distributions, with a $\sim0.4$ dex improved constraint when considering only the IR data (see Table \ref{tab:retrieval_stats}). 

Although IR-only retrievals can obtain improved constraints on the H$_2$O abundance, we continue to consider the STIS data in our analysis of the atmosphere of WASP-17b to gain a more complete understanding of the processes and chemistry present, and to aid in anchoring the abundance constraints. The increased wavelength coverage provided by STIS holds critical information on the role of aerosol scattering, which in turn impacts the understanding of retrieved abundances \citep{Wakeford2018}, and enables us to holistically interpret the atmosphere.

\subsubsection{Comparison with Equilibrium Chemistry} \label{section:atmo}

To better assess the physical nature of our free-chemistry retrievals, we also explore the implications of equilibrium chemistry on the atmosphere of WASP-17b. To do so, we utilised the \verb|ATMO| retrieval code \citep{Amundsen2014, Tremblin2015, Wakeford2017Sci} which has been previously bench-marked against \verb|POSEIDON|'s free chemistry \citep{Lewis2020, Rathcke2021}. \verb|ATMO| computes the chemistry for each atmospheric layer and wavelength such that the opacities are physically self-consistent with the retrieved P-T profile and chemical composition for every likelihood evaluation step. The equilibrium chemical network includes abundances for 175 neutral, 9 ionic, and 93 condensate species. The retrieval considers 22 spectrally active opacity species (see \citealt{Goyal2018} and \citealt{Goyal2020} for a full description and Table \ref{tab:retrieval_stats} for a full list of species), and rainout chemistry such that condensed materials are removed from the gaseous species in layers above their condensation point.  

\begin{table}
\centering
\caption{Statistics and retrieved parameters for the \texttt{ATMO} equilibrium chemistry retrieval. See Table \ref{tab:retrieval_stats} for the full list of spectrally active species.}
\label{tab:atmo_retrieval}
\begin{tabular}{|cc|} 
\hline
                        & 175 neutral, 9 ionic, and      \\
Species                 & 93 condensate species with    \\
                        & 22 spectrally active     \\ 
\hline
Free Parameters         & 6    \\
Degrees of Freedom      & 62    \\
$\chi^2_\nu$            & 1.13    \\
BIC                     & 95.27    \\
$\chi^2$ CDF (\%)       & 77.19   \\ 
\hline
$R_p$ ($R_J$)$\dagger$           & 1.895$^{+0.033}_{-0.013}$      \\
$T$ ($K$)                    & 1250.3$^{+167.7}_{-265.7}$ \\
log($Z/Z_{\odot}$)      & 1.52$^{+0.24}_{-0.57}$ \\
log($X_\mathrm{H_2O}$)$\dagger$  & -2.10$^{+0.22}_{-0.78}$  \\
log($X_\mathrm{CH_4}$)  & -9.64$^{+2.54}_{-1.13}$      \\
log($X_\mathrm{CO}$)$\dagger$    & -1.77$^{+0.23}_{-0.57}$  \\
log($X_\mathrm{CO_2}$)$\dagger$  & -3.93$^{+0.41}_{-0.85}$  \\
log($a$)                & ln($\sigma / \sigma_0$) = 6.34$^{+0.75}_{-2.01}$     \\
$\gamma$                & -4.27$^{+0.52}_{-0.38}$ \\
log($P_\mathrm{cloud}$) & ln($\sigma / \sigma_0$) =       1.27$^{+0.96 }_{-3.89}$      \\
\hline
\end{tabular}
\begin{flushleft}
$\dagger$ Volume mixing ratio at 1 mbar, $R_p$ at 1 mbar.
\end{flushleft}
\end{table}

As with the free-chemistry retrievals, we used an isothermal P-T profile and included the effects from both wavelength-dependent scattering and uniform cloud opacities. In addition, \verb|ATMO| fits for the overall atmospheric metallicity, which can be used to define the abundances for each active species when scaled relative to solar. The results of this retrieval are detailed in Table \ref{tab:atmo_retrieval}. To further evaluate the derived equilibrium chemistry solution, we additionally ran an \verb|ATMO| retrieval wherein the elemental abundances of carbon and oxygen were free parameters along with an overall atmospheric metallicity parameter. We find that carbon and oxygen were fully consistent with each other, with a C/O ratio consistent with solar. 

We show the retrieved transmission spectrum from the \verb|ATMO| equilibrium chemistry retrieval and its 1 and 2$\sigma$ uncertainties in Figure \ref{fig:retrieval}, alongside our highlighted free chemistry retrievals. The equilibrium chemistry model is in good agreement with Models A and C in the NIR, although \verb|ATMO| derives a shallower slope in the optical spectrum where we see the majority of our degeneracies. The equilibrium chemistry retrieval finds similar temperatures and H$_2$O abundances to the free chemistry retrieval high abundance mode (Model C), with a more moderate CO$_2$ abundance, a low CH$_4$ abundance, and a significantly higher CO abundance, resulting in the difference in fit at 4.5\,\micron\, (See lower panels of Figure \ref{fig:retrieval}). Our \verb|ATMO| retrieval finds a $33\times$ solar metallicity atmosphere, in agreement with the super-solar atmosphere found under free chemistry \verb|POSEIDON| configurations where a larger number of spectroscopically active species were considered. 

While the BIC of the \verb|ATMO| retrieval indicates a good fit to the data, given the low number of free parameters required, the somewhat relatively poorer $\chi^2$ CDF demonstrates that equilibrium chemistry is potentially not fully able to explain the measured transmission spectrum of WASP-17b. At the retrieved temperatures, \verb|ATMO| cannot include the opacities of species such as H$^-$, TiO and VO, as seen in free chemistry. Under equilibrium, Ti and V should exist predominantly in solid condensates at temperatures $\lesssim1200$\,$K$\, \citep{Fortney2008}, while the production of significant enough quantities of H$^-$ would require temperatures $\gtrsim$ 2500\,$K$\, \citep{Kitzmann2018}. To this end, it is therefore necessary to consider the potential implications of disequilibrium chemical processes upon the measured atmosphere of WASP-17b.

\subsubsection{Disequilibrium Chemistry} \label{section:diseq}

In this section, we consider the role of disequilibrium chemistry in the interpretation of WASP-17b's measured transmission spectrum. In the context of exoplanetary atmospheres, disequilibrium processes such as photochemistry \citep{Fleury2019}, quenching due to vertical and horizontal mixing \citep{Prinn1977, Cooper2006} and ion chemistry \citep{Lavvas2008}, can allow for the production of molecules at abundances that would not be plausible if the atmosphere were in thermochemical equilibrium. Therefore, the presence or overabundance of certain molecules within an observed transmission spectrum can indicate the presence of such processes. Disequilibrium chemistry is likely to play a key role in governing the chemical make-up of the atmospheres of exoplanets \citep[e.g.,][]{Zahnle2009, Moses2011, Moses2013, Fleury2019, Steinrueck2019, Molliere2020}, and is thus an important consideration for WASP-17b. 

Under equilibrium chemistry and solar metallicity conditions, CO$_2$, CO and H$_2$O are well-mixed throughout the atmosphere \citep{Moses2011}. Furthermore, at the temperatures and metallicities retrieved and the pressures probed for our observations of the atmosphere of WASP-17b, CO should be the dominant carbon-bearing species over CO$_2$, and CH$_4$ would be disfavoured, as the retrieved temperatures would be too high for stable CH$_4$ production (\citealt{Lodders2002}, see also our \verb|ATMO| chemical equilibrium retrieval, Figure \ref{fig:retrieval_stats}). However, across our suite of \verb|POSEIDON| retrievals, we find that the carbon chemistry is dominated by the abundance of CO$_2$, which, along with H$_2$O, shows an increasing abundance trend with CH$_4$. On the other hand, CO is consistently retrieved to a low abundance, albeit with poor constraints. The significant abundance of CO$_2$ compared to CO in the atmosphere could be indicative of a high overall metallicity \citep{HengLyons2016}. However, as the carbon-bearing species in the measured transmission spectrum are encapsulated by the two \textit{Spitzer} photometric points alone, our understanding of the carbon chemistry must remain limited to detections rather than precise abundance constraints (see Section \ref{section:detections}).

Our free chemistry retrieval suite results are further at odds with equilibrium chemistry due to the potential presence of an overabundance of H$^-$ in many retrievals across our suite. Across our free chemistry retrieval suite, the increasing abundance of H$^-$ is driven by the downturn in the final four WFC3/G141 data points, as seen for example in retrievals A and C in Figures \ref{fig:retrieval_stats} and \ref{fig:retrieval}. Here, H$^-$ acts like a uniform opacity source in the optical which turns off at around 1.6\,$\micron$\, coinciding with these data points. Under equilibrium chemistry conditions, the abundances of H$^-$ retrieved in our suite would only be possible at temperatures $\gtrsim$ 2500\,$K$\, \citep{Kitzmann2018}. However, the production of appreciable quantities of H$^-$ could be possible under disequilibrium processes. As described by \citet{Lewis2020}, enhanced e-, H and H$_2$ mixing ratios \citep{Lavvas2014} with production of H$^-$ by H$_2$ dissociative electron attachment and destruction by atomic H collisional detachment, can enable mixing ratios of H$^-$ which are at the orders of magnitude of the retrieved abundances, even for the temperatures expected within the atmosphere of WASP-17b. The conditions required for these processes are likely for hot Jupiters orbiting F-type stars such as WASP-17, making the abundances of H$^-$ seen in our retrieval suite plausible outside of equilibrium conditions. While the downturn at 1.6\,$\micron$, which seems to drive the inclusion of H$^-$, could be explained by patchy clouds \citep[e.g.,][]{Line2016, MacDonald2017}, our explorations of the cloud parameterisation found patchy clouds to be statistically disfavoured for the measured transmission spectrum (see Section \ref{section:bulk}). Additional IR observations around 1.6\,$\micron$\, which overlap with the existing G141 data would help to corroborate the shape of the transmission spectrum in this region, shedding light as to whether H$^-$ is indeed present within the atmosphere of WASP-17b.

\subsubsection{Contextualising Molecular Detections} \label{section:detections}

Table \ref{tab:significances} lists the derived sigma significances \citep[e.g.,][]{Benneke2013} for each retrieval highlighted in Figure \ref{fig:retrieval}, obtained via Bayesian model comparisons \citep[see][]{Trotta2008}.
We confirm the presence of H$_2$O in WASP-17b's atmosphere (to $> 7\,\sigma$ confidence) and report evidence of CO$_2$ absorption ($> 3\,\sigma$), and conclude that the detection of H$_2$O and the inference of CO$_2$ is robust across all our considered retrievals. We also see tentative evidence of CH$_4$ in the A and C retrievals ($\sim 2\,\sigma$). 

Our detection of H$_2$O is largely driven by the precise WFC3/G141 data, as seen for other hot Jupiters with similarly high-quality WFC3 spectra \citep[e.g.][]{Deming2013,Wakeford2018}. We also benefit from the extended wavelength coverage provided by WFC3/G102, which allows us to probe multiple H$_2$O features, similarly seen in the atmospheres of planets such as HAT-P-26b \citep{Wakeford2017Sci}, WASP-39b \citep{Wakeford2018} and WASP-107b \citep{Spake2018}. Our evidence of CO$_2$, meanwhile, arises from the large offset between the two \textit{Spitzer} IRAC photometric points. WASP-17b therefore joins WASP-127b \citep{Spake2021} as one of the only two exoplanets known to exhibit evidence of CO$_2$ absorption. Without the inclusion of optical data (see Section \ref{section:optical}) we see no change in the statistical significance of the detected molecules, confirming that our detection of H$_2$O and CO$_2$ are driven by the WFC3/IR and \textit{Spitzer} data.

While the evidence of carbon species places WASP-17b amongst a rare collection of atmospheres, the lack of IR wavelength coverage currently available limits the properties which can inferred from this evidence, and in the case of WASP-17b, such evidence is also statistically tentative. In the near future, the GTO-1353 observations of WASP-17b with JWST will obtain the complete transmission spectrum of the atmosphere across 0.6--14\,$\micron$, filling in the missing IR coverage. Such a transmission spectrum should be able to confirm the tentative evidence of carbon species, and will likely obtain significantly improved constraints on all atmospheric properties, however the specific predictions are beyond the scope of this work and will be explored in a future study.

\begin{table}
\centering
\caption{Sigma confidence levels of detections of H$_2$O, CO$_2$, and CH$_4$ absorption features for the three \texttt{POSEIDON} retrievals presented in Figure \ref{fig:retrieval}. For full details on species considered in each retrieval, see Table \ref{tab:retrieval_stats}.}
\label{tab:significances}
\begin{tabular}{c|ccc|c}
\hline
 & A & B & C \\ \hline
H$_2$O & 7.90 & 8.19 & 8.17 \\
CO$_2$ & 3.17 & 3.88 & 3.98 \\
CH$_4$ & 2.03 & <1.0 & 2.30 \\
\hline
\end{tabular}
\end{table}

\section{Conclusions} \label{section:discuss_conc}

We have presented a consistent and comprehensive analysis of new HST WFC3/IR and a reanalysis of existing HST STIS and \textit{Spitzer} IRAC observations of the atmosphere of WASP-17b. We take advantage of the implementation of WFC3's spatial scanning mode, along with the many advances in analysis techniques developed in recent years to produce a precise transmission spectrum of WASP-17b from 0.3--5\,\micron. With our spatial scan mode observations, we achieve average uncertainties for WFC3/G141 of 120\,ppm, a marked improvement over previous observations of WASP-17b with this instrument. Across our entire spectrum, we achieve an average uncertainty of 272\,ppm. We also leverage the long temporal baseline provided by TESS observations in order to refine the orbital ephemerides of WASP-17b, resulting in a precise orbital period of 3.73548546\,$\pm$\,0.00000027\,d.

We interpret the measured atmosphere of WASP-17b with an extensive suite of retrieval analyses under both free and equilibrium chemistry using the \verb|POSEIDON| and \verb|ATMO| retrieval codes respectively. The data results in bimodal solutions, with high and low metallicity modes plausible, demonstrating the importance of utilising multiple statistics for model selection.

We find that WASP-17b is best fit by an isothermal P-T profile, consistent with a $\sim 1200\pm200 K$ limb temperature, and requires both a uniform cloud deck and wavelength-dependent scattering aerosol prescription, with a super-Rayleigh gradient. Stellar activity has no discernible impact on the observed transmission spectrum, in agreement with prior observations of the host WASP-17 \citep{Sing2016,Khalafinejad2018}.

We detect absorption due to H$_2$O at $>7\,\sigma$ and find evidence of absorption due to CO$_2$ at $>3\,\sigma$. We find no evidence of absorption due to \ion{Na}{i} or \ion{K}{i} in the atmosphere of WASP-17b, likely as a result of the poorer precision achieved by HST STIS, which was negatively impacted by the partial transits observed by this instrument. Observations of WASP-17b with WFC3/UVIS which utilise the newly refined orbital period could enhance our understanding of its atmosphere, as UVIS has been seen to be able to produce transmission spectra from 0.2--0.8\,\micron\, with superior precision and fewer systematics compared to STIS \citep{Wakeford2020}. 

The retrieval with the best $\chi^2$ CDF results in bimodal posterior distributions for a host of model parameters, and within our full suite of free chemistry retrievals, a wide range of abundances for key species such as H$_2$O, CO$_2$ and CH$_4$ are obtained. Retrievals with and without the inclusion of the STIS optical data indicate that the current available precision at optical wavelengths prevents the retrieval from achieving precise constraints on the abundances of optical species such as Na and K. The optical data improves the constraints on the aerosol scattering properties towards bluer wavelengths, as both the observed spectrum and our interpretation of the data are dependent on these properties. This therefore highlights the need for panchromatic observations when accurately interpreting the transmission spectra of exoplanet atmospheres. 

Our free chemistry retrieval results potentially imply the presence of disequilibrium processes within the atmosphere of WASP-17b. However, given the quality of the optical data, and the lack of coverage in the IR, our limited ability to place constraints on the retrieved abundances of detected molecules mean we can only infer rather than confirm this. A consistent sub-solar C/O ratio is retrieved in all the free chemistry cases, although we are unable place non-model dependent constraints on this due to the limited measurements which probe carbon species. 

Future planned observations with the James Webb Space Telescope as part of GTO-1353 will characterise the transmission spectrum of WASP-17b from 0.6--14\,$\micron$. This additional wavelength coverage, higher resolution and precision will dramatically improve constraints on the carbon inventory of the atmosphere. When combined with the existing transmission spectrum and understanding of the cloud opacity presented in this work, future JWST observations should be able to distinguish between the degenerate solution we currently obtain, and will enable constraints to be placed upon key atmospheric chemical species that will help to trace the formation and evolution of WASP-17b and the processes shaping its atmosphere today.

\section*{Acknowledgements}

We thank the anonymous referee for their helpful suggestions and comments. We also thank T.J. Wilson for useful discussions on the statistics of retrieval comparison. This work is based on observations made with the NASA/ESA Hubble Space Telescope, HST-GO-14918 and GO-12473, that were obtained at the Space Telescope Science Institute, which is operated by the Association of Universities for Research in Astronomy, Inc. We thank J-M Desert for use of data from their \textit{Spitzer} program 90092. L. Alderson acknowledges funding from STFC grant ST/W507337/1 and from the University of Bristol School of Physics PhD Scholarship Fund. N.K.L and R.J.M acknowledge support from NASA grant 80NSSC20K0586 issued through the James Webb Space Telescope (JWST) Guaranteed Time Observer Program to the JWST Telescope Scientist Team. 

Author contributions: L. Alderson led the data analysis, interpretation, and write-up of this study. 
H.R. Wakeford conducted the STIS analysis.
E. M. May conducted the analysis of the Spitzer IRAC data. 
L. Alderson and D. Grant conducted the analysis on the TESS data and fitting of the orbital period with insight from K. Stevenson.
L. Alderson, R. MacDonald, H.R. Wakeford, and D.K. Sing conducted the retrieval analysis.
L. Alderson and J. Fowler developed the data extraction pipeline.
J. Goyal consulted on the use of forward models used in this study.
H.R. Wakeford conceived and proposed the study along with, N.K. Lewis, D.K. Sing, N.E. Batalha, M. Clampin and T. Kataria. 
H.R. Wakeford and N.K. Lewis advised throughout the study.
All authors read and approved the manuscript.

\section*{Data Availability}
All HST and TESS data are available through MAST\footnote{\href{https://exo.mast.stsci.edu/exomast\_planet.html?planet=wasp17b}{exo.MAST.STScI/WASP-17b}}. WFC3 G102 and G141 data can be found on MAST as part of HST GO-14918 (PI. H.R. Wakeford) visits 1 and 32. Please note the pointing failed during visit 2 of this program resulting in unusable data. 
STIS/G430L and G750L data can be found on MAST as part of GO-12473 (PI. D.K. Sing) visits 5, 6, and 19. WASP-17 was measured with TESS in sector 12 and 38 with corrected light curves available through MAST.
Spitzer IRAC data is part of program 90092 (PI J-M, Desert) and can be found through IPAC.

\section*{Software Used}
This analysis made use of components of the IDL Astronomy Users Library \citep{IDL} and the Python packages: NumPy \citep{numpy}, SciPy \citep{scipy}, MatPlotLib \citep{matplotlib}, AstroPy \citep{astropy}, and Photutils \citep{photutils}, Batman \citep{Kreidberg2015}, exoctk \citep{exoctk2021}, lightkurve \citep{Lightkurve2018}, emcee \citep{Foreman-Mackey2013}, celerite \citep{Foreman-Mackey2017,Foreman-Mackey2018}, PyMultinest \citep{pymultinest}. This research made use of ExoTiC-ISM \citep{Laginja2020}, a software package for marginalised transit parameters, which was developed based on the work by \cite{Wakeford2016}. 
 


\bibliographystyle{mnras}
\bibliography{refs}



\appendix

\section{Additional Light Curve Figures}

In this appendix we show the individual light curves from each of our datasets and the associated spectroscopic channels. 

\begin{figure*}
    \centering
    \includegraphics[width=\linewidth]{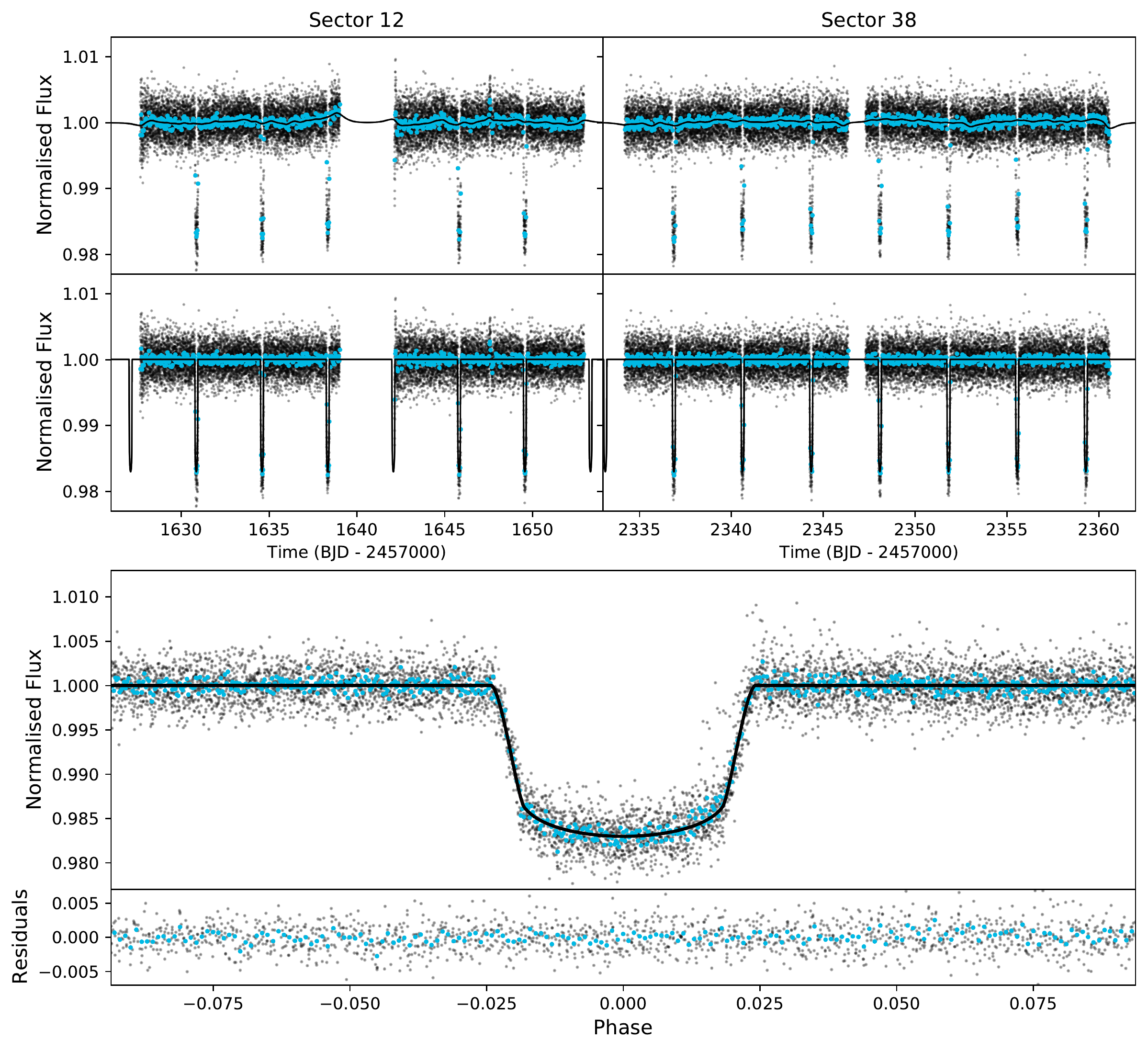}
    \caption{Transit light curves of WASP-17b obtained by TESS. Top: Raw light curves for Sector 12 (left) and Sector 38 (right). The unbinned flux is shown by the black data points while the binned flux is shown in blue. The systematic model is given by the black line. Middle: Detrended light curves for Sector 12 (left) and Sector 38 (right). The unbinned flux is shown by the black data points while the binned flux is shown in blue. The transit model is given by the black line. Bottom: Phase folded transit light curve of WASP-17b for both Sectors 12 and 38 along with residuals.}
    \label{fig:tess_folded_lc}
\end{figure*}

\begin{figure*}
    \centering
    \includegraphics[width=\linewidth]{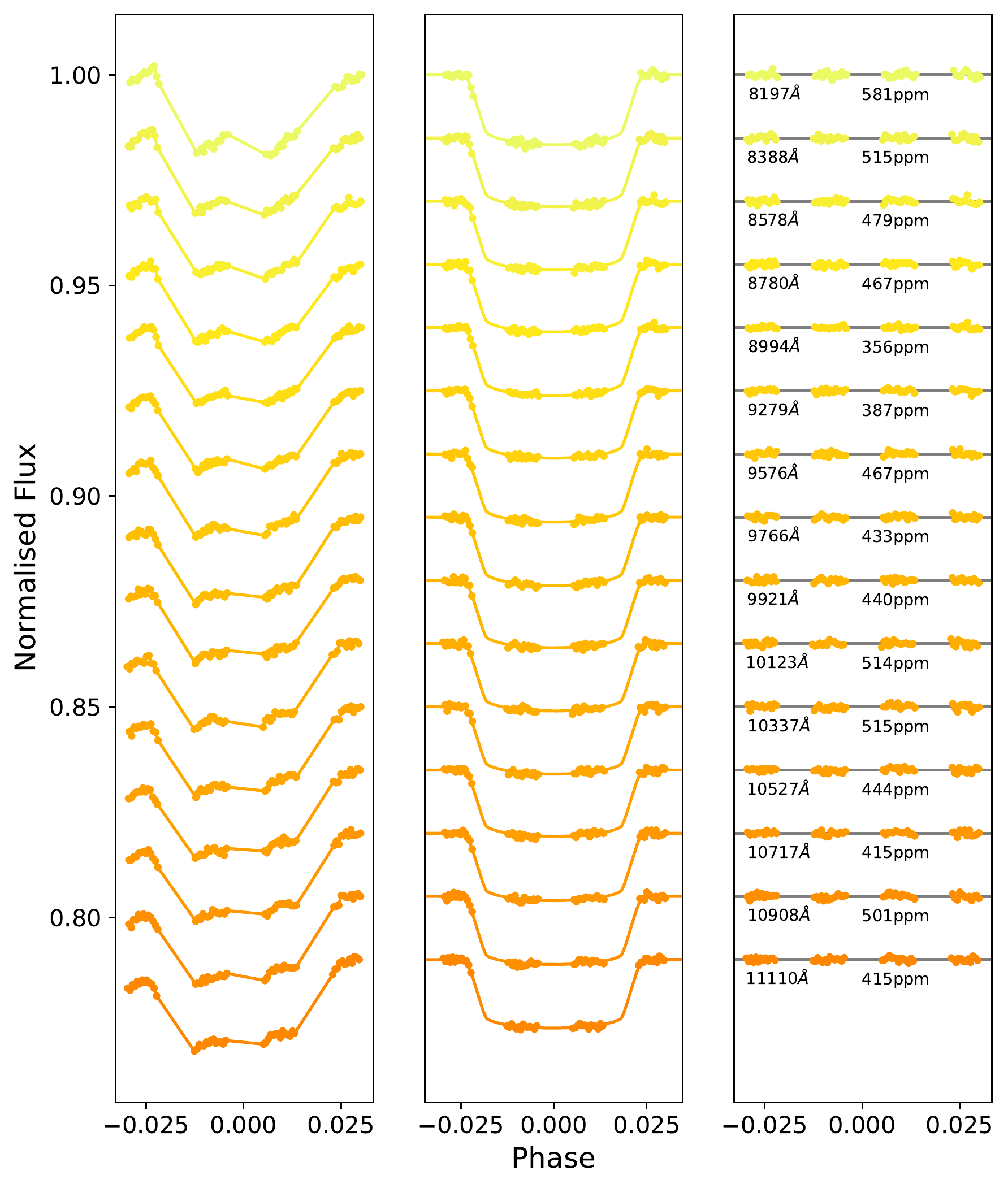}
    \caption{Spectroscopic light curves of WASP-17b from WFC3/G102 visit 32. Left: Raw light curves. Middle: Corrected light curves and best fit model obtained with an ExoTiC-ISM routine. Right: Corrected light curve residuals. The central wavelength and standard deviation of the residuals for each wavelength bin are shown below the corresponding residuals.}
    \label{fig:g102_slcs}
\end{figure*}

\begin{figure*}
    \centering
    \includegraphics[width=\linewidth]{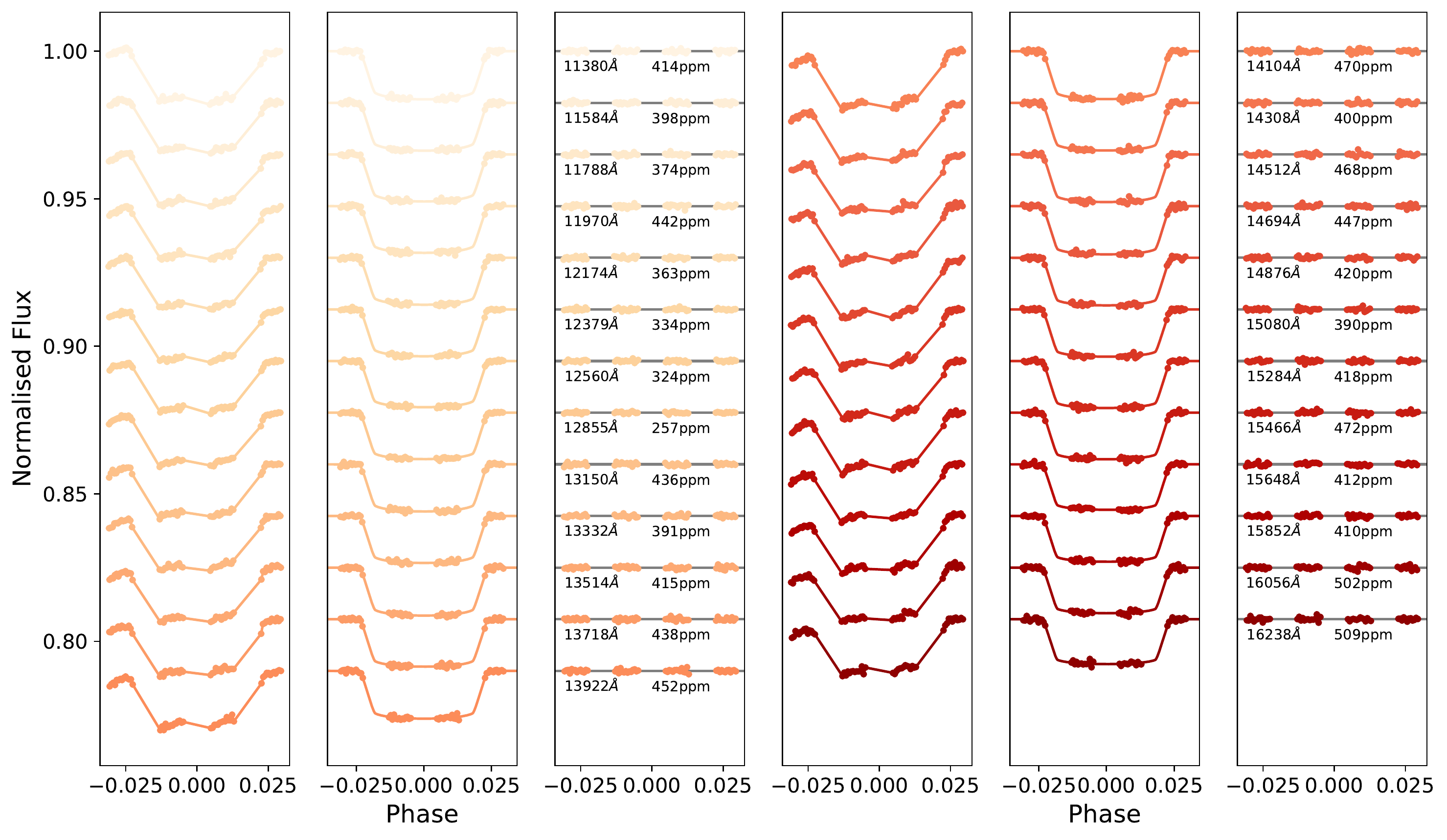}
    \caption{Spectroscopic light curves of WASP-17b from WFC3/G141 visit 1. For details see Figure \ref{fig:g102_slcs}.}
    \label{fig:g141_slcs}
\end{figure*}

\begin{figure*}
    \centering
    \includegraphics[width=0.49\linewidth]{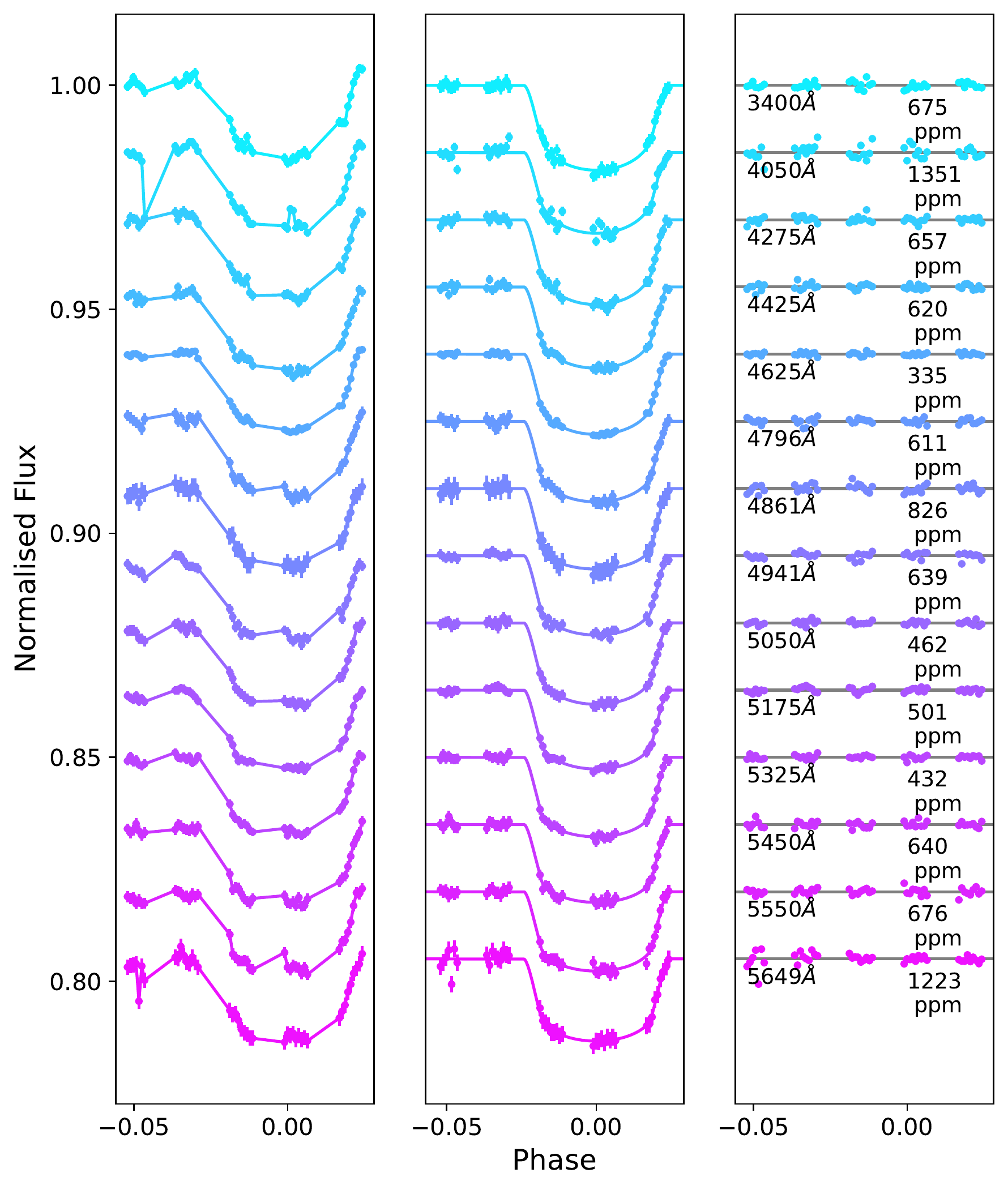}
    \includegraphics[width=0.49\linewidth]{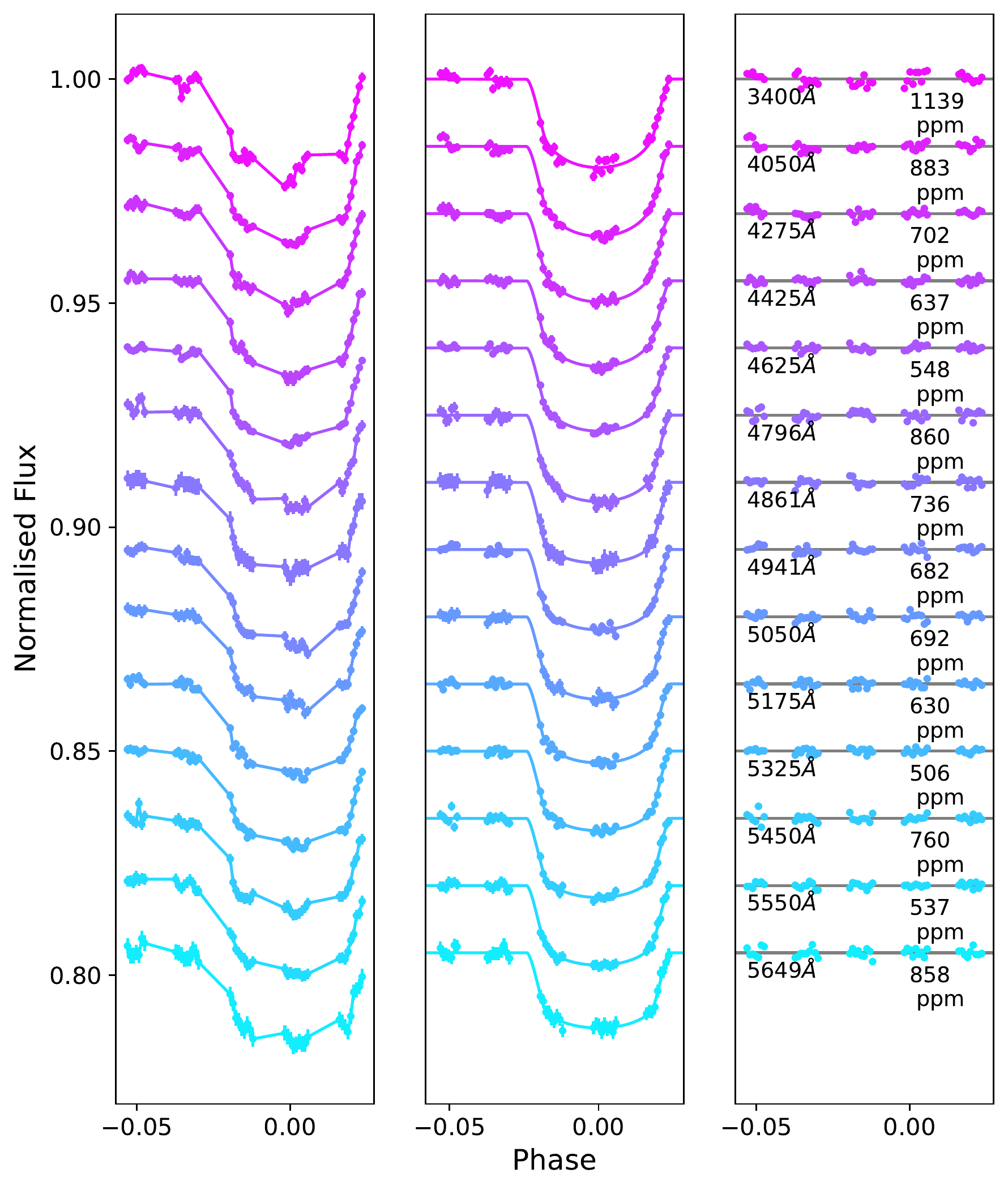}
    \caption{Spectroscopic light curves of WASP-17b from STIS/G430L visit 5 (left) and visit 6 (right). For details see Figure \ref{fig:g102_slcs}.}
    \label{fig:g430_slcs}
\end{figure*}

\begin{figure*}
    \centering
    \includegraphics[width=\linewidth]{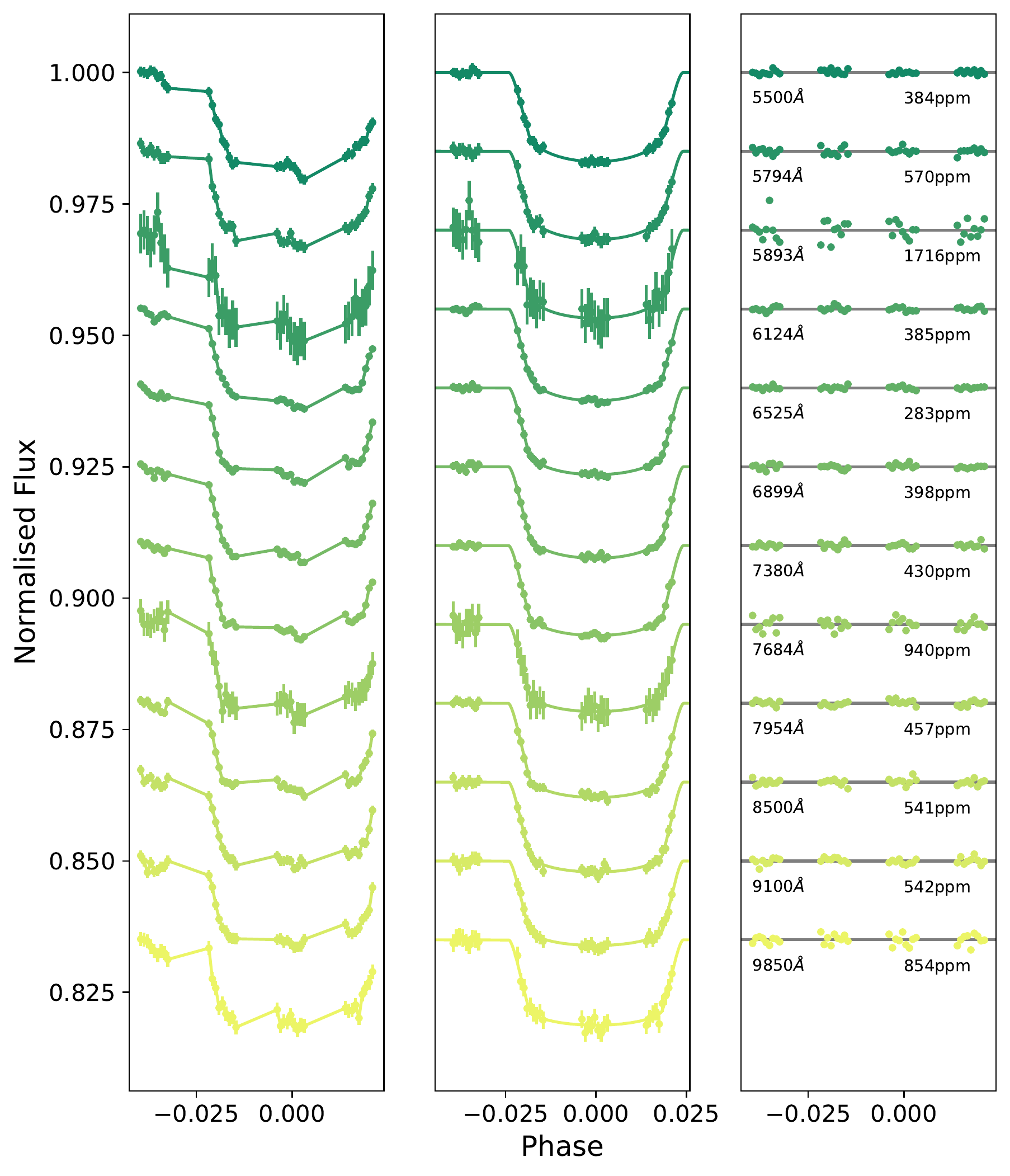}
    \caption{Spectroscopic light curves of WASP-17b from STIS/G750L visit 19. For details see Figure \ref{fig:g102_slcs}.}
    \label{fig:g750_slcs}
\end{figure*}


\bsp	
\label{lastpage}
\end{document}